\documentclass[aps,pre,superscriptaddress]{revtex4}

\usepackage[dvips]{graphicx}
\usepackage{mathrsfs}
\usepackage{amsmath,amssymb}
\usepackage{hhline}
\usepackage{bm}% bold math

\usepackage[dvips]{color}
\usepackage{comment}

\newcommand{\red}[1]{}

\newcommand{\be}{\begin{equation}}
\newcommand{\ee}{\end{equation}}
\newcommand{\bi}{\begin{itemize}}
\newcommand{\ei}{\end{itemize}}

\begin{document}

\title{Impact of assortative mixing by mask-wearing on the propagation of epidemics in networks}

\author{Hiromu Watanabe}
\affiliation{Graduate School of Science and Engineering, Ibaraki University, 2-1-1, Bunkyo, Mito, 310-8512, Japan}
\author{Takehisa Hasegawa}
\email{takehisa.hasegawa.sci@vc.ibaraki.ac.jp}
\affiliation{Graduate School of Science and Engineering, Ibaraki University, 2-1-1, Bunkyo, Mito, 310-8512, Japan}

\begin{abstract} 
In this study, we discuss the impacts of assortative mixing by mask-wearing on the effectiveness of mask use in suppressing the propagation of epidemics. 
We employ the mask model, which is an epidemic model involving mask wearers and non-mask wearers. 
We derive the occurrence probability and mean size of large outbreaks, epidemic threshold, and average epidemic size for the mask model in an assortatively mixed random network that follows an arbitrary degree distribution.
Applying our analysis to the Poisson random networks, we find that the assortative (disassortative) mixing by mask-wearing decreases (increases) the epidemic threshold. 
Assortative mixing, the tendency for (non-)mask wearers to prefer to connect with (non-)mask wearers, is not effective in containing epidemics in that the transmissibility required for large outbreaks to occur is small.
On the other hand, in high-transmissibility cases, mask use is most effective in decreasing the occurrence probability and mean size of large outbreaks, as well as the average epidemic size, when the mixing pattern is strongly assortative. 
Strongly assortative mixing, resulting in the separation of mask wearers and non-mask wearers, reduces the probability and degree of a large outbreak in high-transmissibility cases, although it allows a large outbreak to occur even in low-transmissibility cases. 
In scale-free networks, mask use is most effective when the mixing pattern is strongly assortative and when it is maximally disassortative, provided that the mask coverage is not low.
Both for the Poisson random and scale-free networks, all analytical treatments are in good agreement with simulation results.
\end{abstract}

\maketitle

%%%%%%%%%%%%%%%%%%%%%%%%%%%%%%%%%%%

\section{Introduction}

Since its first outbreak in Wuhan (China) in December 2019, the coronavirus disease (COVID-19), caused by severe acute respiratory syndrome coronavirus 2 (SARS-CoV-2), has spread rapidly globally.
On March 2, 2020, the World Health Organization (WHO) declared COVID-19 a pandemic~\footnote{https://www.who.int/director-general/speeches/detail/who-director-general-s-opening-remarks-at-the-media-briefing-on-covid-19---11-march-2020}. 
Even in 2022, the COVID-19 pandemic shows no signs of cessation.
In response to the COVID-19 pandemic, various public health and social measures have been implemented globally~\cite{estrada2020covid}. 
The large-scale preventative measures implemented, such as the imposition of travel restrictions, closure of schools and businesses, prohibition of group gatherings, and implementation of lockdowns, have severely disrupted social and economic activities. 
To limit the spread of the virus, washing of hands, social distancing, and mask usage have become the new prerequisites for partaking in social activities. 
Thus, it is important to investigate the effectiveness of mask use in suppressing the virus transmission in various situations.

To understand the spread of diseases beyond the population level, it is necessary to consider a contact network of individuals~\cite{barabasi2016network,newman2018networks}. 
Numerous studies have shown that the structure of an underlying contact network significantly impacts the propagation of epidemics (see review~\cite{pastor2015epidemic} and references therein).
For instance, an epidemic can propagate even with a small infection rate if the underlying network is highly heterogeneous~\cite{pastor2001epidemic,moreno2002epidemic}. 
It should further be noted that population heterogeneity in terms of socio-demographic and health factors also affects disease dynamics~\cite{markovivc2021socio}.
Network scientists have explored network-based strategies to efficiently suppress epidemics, including optimal vaccine allocation~\cite{gallos2007improving, wang2016statistical}, rapid isolation of infected individuals~\cite{takaguchi2014suppressing,hasegawa2017efficiency,strona2018rapid}, social distancing~\cite{gutin2021effect,maharaj2012controlling,valdez2012intermittent,maheshwari2020network,nande2021dynamics}, and community lockdown~\cite{gosak2021community}. 
Recent studies (since COVID-19) have discussed the effectiveness of mask use in limiting the spread of the virus in networks~\cite{li2020mask,eikenberry2020mask,leung2020contact,yagan2021modeling,tian2021analysis,ying2021modelling,sridhar2021leveraging,lee2021epidemic,tian2021role}.
Ya{\u{g}}an et al.~\cite{yagan2021modeling} proposed the implementation of a susceptible-infected-recovered (SIR) model in a network consisting of mask wearers and non-mask wearers, based on a multiple-strain model, where a virus may mutate into a different strain~\cite{alexander2010risk,eletreby2020effects}. 
They referred to it as the {\it mask model}. 
Tian et al. derived the expected size and occurrence probability of large outbreaks in the mask model~\cite{tian2021analysis}. 
Additionally, Lee and Zhu considered the same model to estimate the infection and recovery rates from John Hopkins University's COVID-19 database. 
They further assessed the mask coverage using the maximum likelihood estimation method~\cite{lee2021epidemic}. 

Homophily, the tendency for individuals to connect with similar others, has been found in a wide variety of attributes such as race, age, and gender~\cite{mcpherson2001birds}. 
An experimental study by Centola~\cite{centola2011experimental} discerned homophily in the adoption of health behaviors.
Since COVID-19, some studies have reported homophily in mask use~\cite{haischer2020wearing,woodcock2021role}.
Despite the guidelines from the United States Centers for Disease Control and Prevention~\footnote{https://www.cdc.gov/coronavirus/2019-ncov/prevent-getting-sick/diy-cloth-face-coverings.html} and the WHO~\footnote{https://www.who.int/emergencies/diseases/novel-coronavirus-2019/advice-for-public/when-and-how-to-use-masks}, some members of the population do not wear masks, or even if they do, they do not wear highly protective masks, such as surgical masks and N95 masks. 
Homophily in mask use will be factored in as {\it assortative mixing}~\cite{newman2003mixing} in networks.
Assortative mixing here means a tendency for (non-)mask wearers to prefer to connect with other (non-)mask wearers and disassortative mixing means a tendency for mask-wearers to prefer to connect with non-mask wearers.
Whereas previous studies~\cite{yagan2021modeling,tian2021analysis,sridhar2021leveraging} assumed that mask wearers and non-mask wearers are randomly distributed, the present study considers mixing pattern by mask-wearing to discuss the impact of assortative mixing on the effectiveness of mask use in suppressing the propagation of epidemics.

In this paper, we analyze the mask model in assortatively mixed random networks that follow an arbitrary degree distribution. 
We derive the occurrence probability of large outbreaks, mean size of large outbreaks, and average epidemic size of the mask model in assortatively mixed random networks using probability generating functions.
Applying our analysis to the mask model in Poisson random networks, we establish that the assortative (disassortative) mixing by mask-wearing decreases (increases) the epidemic threshold. 
It means that disassortative mixing is effective in containing epidemics in that a high transmissibility is required for large outbreaks.
On the other hand, the result for the case of high transmissibility shows that the mask use is most effective in reducing the probability and mean size of large outbreaks when the mixing pattern is strongly assortative.
The separation of mask wearers and non-mask wearers, which occurs in strongly assortative networks, reduces the probability and degree of a large outbreak in high-transmissibility cases, although it allows a large outbreak to occur even in low-transmissibility cases. 
For scale-free networks, we establish that mask use is most effective when the mixing pattern is strongly assortative and maximally disassortative (provided that the mask coverage is not extremely low).
We also performed extensive simulations to confirm that our analytical results are in full agreement with the simulation results. 

The paper is organized as follows. 
In Sec.~\ref{sec:model}, we introduce the mask model, which is a generalization of a discrete-time SIR model, and assortatively mixed random networks proposed in~\cite{newman2003mixing}. 
In Sec.~\ref{sec:analysis}, we derive the occurrence probability and mean size of large outbreaks, as well as the average epidemic size, for the mask model in assortatively mixed random networks with an arbitrary degree distribution and mixing matrix. 
In Sec.~\ref{sec:results}, we discuss our results for the mask model in two representative networks: a Poisson network and a scale-free network. 
A comparison between the analytical treatment and simulation results is shown. 
Section~\ref{sec:summary} is devoted to the summary. 

\section{Model \label{sec:model}}

In this study, we treat a discrete-time SIR-type model as follows~\cite{newman2002spread}.
Let us consider a network of $n$ nodes and edges, where the nodes represent individuals and the edges represent contact between them.
We assume that each node can be in one of three states: susceptible, infected, or recovered.
The transmission of infections to a susceptible node occurs, with a constant probability, when the latter is adjacent to an infected node.
The system evolves in discrete times as follows.
The dynamics start with an initial state where a single seed node is randomly selected and is set to infected, while the other nodes are set to susceptible.
The seed node infects each of its adjacent nodes independently with probability $T$. 
Here $T$ is called the {\it transmissibility} of the disease~\cite{newman2002spread}. 
Thereafter, the seed node spontaneously recovers (recovered node), i.e., it can neither infect nor be infected. 
In a next step, a newly infected node infects each of its adjacent susceptible nodes with probability $T$ and subsequently recovers.
This process continues until no newly infected node appears (or equivalently, only susceptible and recovered nodes exist in the network). 

The scale of outbreaks in a network depends entirely on whether the transmissibility, $T$, exceeds the epidemic threshold, $T_c$, or not. 
For $T<T_c$, epidemics starting from a single seed node always terminate at an early stage and the fraction of nodes eventually infected is negligibly small. 
For $T>T_c$, a single infected seed node can cause a large outbreak (an outbreak of finite relative size in a large network) with nonzero probability. 
The average epidemic size $s_{\rm av}$, defined as the average fraction of recovered nodes in the final state, is employed to characterize this phase transition: in the limit $n \to \infty$, $s_{\rm av}=0$ when $T \le T_c$ and $s_{\rm av}>0$ when $T>T_c$. 
It is known~\cite{newman2002spread} that the epidemic threshold of the SIR model in random networks with degree distribution $p(k)$ is 
\be
T_c = \frac{\langle k \rangle}{\langle k^2 \rangle -\langle k \rangle}, \label{eq:knownThreshold}
\ee
where $\langle A(k) \rangle = \sum_k A(k) p(k)$.

The mask model was proposed as a variant of the above epidemic model, where some individuals wear masks, while others do not~\cite{yagan2021modeling,tian2021analysis}. 
Each node in the mask model is classified into one of the following two types: type-1 nodes representing mask wearers and type-2 nodes representing non-mask wearers. 
We denote the mask coverage by $p_1$ and assume that the $p_1$ fraction of randomly chosen nodes is of type 1, while the other $p_2(=1-p_1)$ fraction is of type 2. 
In the mask model, there are four types of contacts between infected and susceptible nodes, depending on whether each of them wears a mask or not. 
We designate $T_{xy}$ as the transmissibility from an infected node of type $x$ ($x=1,2$) to a susceptible node of type $y$ ($y=1,2$). 
As the previous study~\cite{yagan2021modeling} pointed out, mask use is effective in reducing the droplets expelled from an infected individual, but it is less effective in preventing an individual from inhaling droplets from an infected non-mask wearer~\cite{brienen2010effect, lai2012effectiveness, davies2013testing,ueki2020effectiveness}; thus, we suppose that $T_{11}<T_{12}<T_{21}<T_{22}$. 

The objective of this study is to assess the impact of assortative mixing by mask-wearing on the effectiveness of mask use in suppressing the propagation of epidemics. 
The mixing pattern is characterized by mixing matrix ${\bm e}$, whose element $e_{xy}$ denotes the fraction of edges connecting a type-$x$ node and a type-$y$ node~\cite{newman2003mixing}. 
For a network in which nodes can be one of two types (wearing or non-wearing group), ${\bm e}=(e_{xy})$ satisfies: 
\be
e_{xy}=e_{yx},\; p_x = \sum_{y=1,2}e_{xy}=\sum_{y=1,2}e_{yx}, \; \sum_{x,y=1,2} e_{xy} = 1.
\ee
The tendency of (non-)mask wearers to be connected to other (non-)mask wearers is quantified by the {\it assortativity coefficient} $r$~\cite{newman2003mixing}. 
For the present case, the assortativity coefficient of mask-wearing is determined only by $p_1$ ($0 < p_1 < 1$) and $e_{12}$ ($0 \le e_{12} \le {\rm min}\{p_1,1-p_1\}$): 
\be
r
=\frac{\sum_{x} e_{xx} - \sum_x p_x^2}{1-\sum_x p_x^2}
=\frac{p_1(1-p_1) -e_{12}}{p_1(1-p_1)},
\ee
where the range of $r$ is
\be
r_{\rm min} \le r \le 1,
\ee
$r_{\rm min}$ being
\be
r_{\rm min} = \begin{cases}
-p_1/(1-p_1) & (0 < p_1 \le 1/2) \\
-(1-p_1)/p_1 & (1/2 \le p_1 < 1)
\end{cases}. 
\label{eq:rmin}
\ee
The network is {\it assortative}, meaning that mask-wearers are likely to be connected to other mask-wearers, {\it neutral}, and {\it disassortative}, meaning that mask-wearers are unlikely to be connected to other mask-wearers, when $r>0$, $r \approx 0$, and $r<0$, respectively. 

We introduce a procedure~\cite{newman2003mixing} for generating assortatively mixed and random networks with an arbitrary degree distribution $p(k)$ and mixing matrix ${\bm e}$. 
Let the number of edges $m$, degree distribution $p(k)$, and mixing matrix ${\bm e}$ be predetermined. 
The procedure is as follows: 
(i) we prepare $m$ edges so that the type of ends of each edge follow ${\bm e}$. 
Later, $m e_{11}$ edges connect two type-$1$ nodes, $m(e_{12}+e_{21})$ edges connect a type-$1$ node and a type-$2$ node, and $m e_{22}$ edges connect two type-$2$ nodes. 
The total degrees of type-1 nodes, $m_{1}^{(\rm e)}$, and type-2 nodes, $m_{2}^{(\rm e)}$, are expressed as $m_{1}^{(\rm e)}=m (2 e_{11} + e_{12})$ and $m_{2}^{(\rm e)}=m (e_{21} + 2 e_{22})$, respectively.
Subsequently, we determine the number, $n_x$, of type-$x$ nodes by $n_x=m_x^{(\rm e)}/\langle k \rangle$ ($x=1,2$).
(ii) 
For each type $x$, we assign $k_i$ stubs to each node $i$ according to the distribution $p(k)$. 
We select a type-$x$ node at random and reassign stubs to it from $p(k)$, repeating the process until the total degree of type-$x$ nodes is equal to $m_x^{(\rm e)}$. 
(iii) 
We repeatedly join a randomly selected pair of stubs to form an edge, thereby realizing a network.
Here, multiple edges and self-loops are prohibited.
By implementing this procedure while tuning $p_1$ and $e_{12}$, we obtain assortatively mixed random networks with a given distribution $p(k)$ and level of assortative mixing $r$.

\begin{figure}[tb]
 \centering
  \includegraphics[width=12cm]{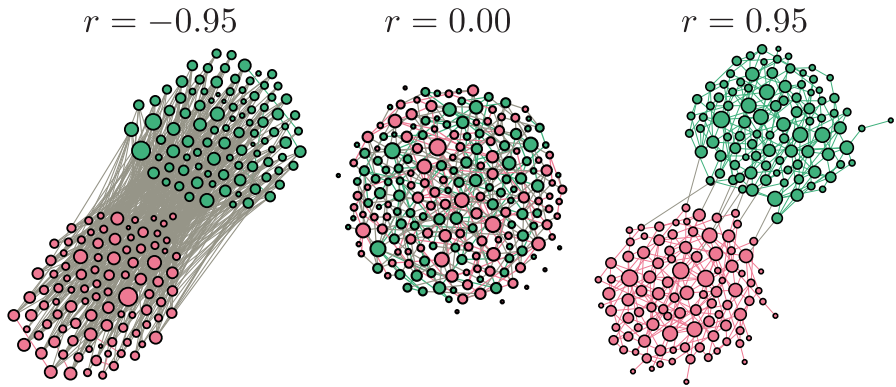}
 \caption{
Illustration of a disassortative network ($r = -0.95$), a neutral network ($r = 0.00$), and an assortative network ($r = 0.95$), from left to right. 
The mask coverage $p_1$ is set to $p_1=0.5$.
All networks have $n \approx 200$ nodes and $m = 400$ edges, and follow a Poisson degree distribution with $\langle k \rangle \approx 4$.
The red nodes and green nodes represent mask wearers and non-mask wearers, respectively.
The more $r$ is negative, the number of edges connecting a mask wearer and a non-mask wearer increases and the network becomes closer to bipartite. 
The more positive $r$ is, the number of edges connecting a mask wearer and a non-mask wearer decreases and the separation of groups of mask-wearers and non-mask-wearers becomes more pronounced.
 }
 \label{fig:Snapshot}
 \end{figure}

Figure~\ref{fig:Snapshot} illustrates assortatively mixed random networks with $r = -0.95$, $0.0$, and $0.95$. 
The more $r$ is negative, the number of edges connecting a mask wearer and a non-mask wearer increases and the network becomes closer to bipartite. 
The more positive $r$ is, the number of edges connecting a mask wearer and a non-mask wearer decreases and the separation of groups of mask-wearers and non-mask-wearers becomes more pronounced.
In particular, the perfectly disassortative case ($r=-1$) results in a bipartite network of mask wearers and non-mask wearers, and the perfectly assortative case ($r=1$) results in two disconnected random networks, one consisting of mask wearers and the other consisting of non-mask wearers.
The case of $r=0$ corresponds to the mask model on random networks analyzed in~\cite{yagan2021modeling,tian2021analysis}.

%%%%%%%%%%%%%%%%%%%%%%%%%%%%%%%%%%%%%
%%%%%%%%%%%%%%%%%%%%%%%%%%%%%%%%%%%%%

\section{Analysis \label{sec:analysis}}

The mask model can be mapped onto the heterogeneous bond percolation of a multitype network~\cite{allard2009heterogeneous}. 
In Ref.~\cite{allard2009heterogeneous}, Allard et al. derived a number of percolation properties for heterogeneous bond percolation on multitype networks with arbitrary degree distribution, using a generating function formalism.
In the context of epidemiology, their generating functions yield the expected size of small outbreaks, the mean size of large outbreaks, the occurrence probability of large outbreaks, and the critical transmissibility set, over which a seed node leads to a large outbreak with nonzero probability.
Similar to Ref.~\cite{allard2009heterogeneous}, we obtain the relative size of the giant component, $s_{\rm L}$, of heterogeneous bond percolation on assortatively mixed random networks.
This giant component size $s_{\rm L}$ corresponds to the mean size of large outbreaks. 
Additionally, we derive the occurrence probability of large outbreaks $P_{\rm L}$ using the probability generating functions. 
Thereafter, the average epidemic size $s_{\rm av}$ is given by $P_{\rm L}$ and $s_{\rm L}$. 
It should be noted that in heterogeneous bond percolation, the mean size of large outbreaks $s_{\rm L}$, given as the probability that a randomly chosen node is reached from an infinite connected component does not always coincide with the occurrence probability $P_{\rm L}$, given as the probability that a randomly chosen node does not link only to finite components~\cite{allard2009heterogeneous}. 
The difference between $s_{\rm L}$ and $P_{\rm L}$ has been known for the SIR model on semidirected networks~\cite{meyers2006predicting} and the SIR model with heterogeneity of infectious periods, infectivity, and susceptibility~\cite{kenah2007second, miller2007epidemic} prior to~\cite{allard2009heterogeneous}.

Let us consider an infinitely large, assortatively mixed, and random network with degree distribution $p(k)$ and mixing matrix ${\bm e}$. 
A node reached by following a randomly chosen edge (or, a randomly chosen neighbor of a node in a random network) has $k$ other edges with probability $q(k)=(k+1)p(k+1)/\langle k \rangle$, where $q(k)$ represents the excess degree distribution. 
The generating functions for $p(k)$ and $q(k)$ are defined as 
\be
G_p(x)=\sum_k p(k) x^k 
\quad {\rm and} \quad
G_q(x)=\sum_k q(k) x^k,
\ee
respectively. 

Now, we consider the heterogeneous bond percolation~\cite{allard2009heterogeneous} with probability $\{T_{xy}\}$ ($T_{xy}$ is the occupation probability of an edge from a type-$x$ node to a type-$y$ node). 
We denote by $u_x$ the probability that a type-$x$ node ($x=1,2$) does not receive the infection (occupied edge) from a neighbor in the giant component.
The probability, $p_{xy}$, that a randomly chosen neighbor of a type-$x$ node is of type $y$ is given by $p_{xy} = e_{xy}/p_x$.
Noting that a type-$y$ neighbor belongs to the giant component is $1-\sum_k q(k) u_y^k = 1 - G_q(u_y)$, the probability that a neighbor of a type-$x$ node is of type $y$, belongs to the giant component, and infects the type-$x$ node is given by $p_{xy} T_{yx} (1-G_q(u_y))$. 
Thus, in the infinitely large, assortatively mixed, random network with ${\bm e}$, $u_1$ and $u_2$ obey the following equations: 
\begin{subequations}
\begin{align}
u_1 &= 1 - p_{11} T_{11}(1-G_q (u_1)) - p_{12} T_{21}(1-G_q (u_2)), \label{eq:u1}\\
u_2 &= 1 - p_{21} T_{12}(1-G_q (u_1)) - p_{22} T_{22}(1-G_q (u_2)). \label{eq:u2}
\end{align} \label{eq:u}
\end{subequations}
A randomly chosen node is of type $x$ with probability $p_x$, and it does not belong to the giant component with probability $\sum_k p(k) u_x^k = G_p(u_x)$. 
The relative size of the giant component $s_{\rm L}$ is one minus the probability that a randomly chosen node is not connected to the giant component:
\be
s_{\rm L} = 1- p_1 G_p(u_1) - p_2 G_p(u_2), \label{eq:S}
\ee
where $u_1$ and $u_2$ are given as the solutions of Eq.~(\ref{eq:u}).
Since this giant component can be regarded as the set of nodes that will eventually be infected in an epidemic if a large outbreak occurs, $s_{\rm L}$ corresponds to the mean size of large outbreaks (the expected size of an outbreak conditioned on it being a large one). 

Furthermore, we derive the critical transmissibility set from Eq.~(\ref{eq:u}). 
For this, we consider the stability of the trivial solution of Eq.~(\ref{eq:u}), i.e., $(u_1, u_2)=(1,1)$, assuming that $u_1 \approx 1-\epsilon_1$ and $u_2 \approx 1-\epsilon_2$. 
By expanding Eq.~(\ref{eq:u}) to the leading order in $\epsilon_1$ and $\epsilon_2$, we have
\be
{\bm \epsilon} = A {\bm \epsilon},
\ee
where
\be
{\bm \epsilon}=
\begin{pmatrix}
\epsilon_1 \\
\epsilon_2
\end{pmatrix}
\quad {\rm and} \quad
A=
G_q' (1)
\begin{pmatrix}
p_{11} T_{11} & p_{12} T_{21} \\
p_{21} T_{12} & p_{22} T_{22}
\end{pmatrix}.
\ee
Thus, the critical transmissibility set is obtained from ${\rm det} (A-I)=0$; more specifically,
\be
(p_{11}T_{11} G_q'(1) -1)(p_{22} T_{22} G_q'(1) -1)-p_{12} p_{21} T_{21}T_{12} G_q'(1)^2 =0. \label{eq:thresholdCondition}
\ee
It should be noted that condition (\ref{eq:thresholdCondition}) reduces to the epidemic threshold (\ref{eq:knownThreshold}) for the SIR model with transmissibility $T$ when mask usage is completely ineffective, i.e., $T_{11}=T_{12}=T_{21}=T_{22}=T$.

The probability $p_{xy}$ is written in terms of $p_1$ and $r$ as $p_{11} = 1-(1-p_1)(1-r)$, $p_{12}=(1-p_1)(1-r)$, $p_{21}=p_1(1-r)$, and $p_{22}=1-p_1(1-r)$.
Equation~(\ref{eq:thresholdCondition}) is then rewritten as
\be
((r+p_1(1-r)) T_{11} G_q'(1) -1)((1-p_1(1-r))T_{22} G_q'(1) -1) - p_1(1-p_1)(1-r)^2 G_q'(1)^2 T_{12} T_{21} =0. \label{eq:thresholdCondition2}
\ee
We should mention the following cases.
(i) For the neutral case ($r=0$), Eq.~(\ref{eq:thresholdCondition2}) reduces to
\be
(p_1 T_{11} G_q'(1)-1)((1-p_1) T_{22} G_q'(1)-1) - p_1(1-p_1) T_{12} T_{21} G_q'(1)^2 =0,
\ee
which is identical to that of the mask model in random networks already reported in~\cite{tian2021analysis}.
(ii) For the perfectly assortative case ($r=1$), Eq.~(\ref{eq:thresholdCondition2}) reduces to
\be
(T_{11} G_q'(1)-1)(T_{22} G_q'(1)-1) = 0. \label{eq:thresholdAssortative}
\ee
In this case, the critical transmissibility set is not affected by $T_{12}$ and $T_{21}$ because no edges exist between mask wearers and non-mask wearers.
Supposing that $T_{11}<T_{22}$ (i.e., mask has an effect in reducing the transmissibility), the epidemic threshold is given by $T_{22} G_q'(1)=0$.
(iii) For the maximally disassortative case ($r=r_{\rm min}$), Eq.~(\ref{eq:thresholdCondition2}) reduces to
\be
\begin{cases}
(1-2p_1) T_{22} G_q'(1) - p_1 T_{12}T_{21} G_q'(1)^2 = 1-p_1 & (0<p_1<1/2) \\
(2p_1-1) T_{11} G_q'(1) +(1-p_1) T_{12}T_{21} G_q'(1)^2 = p_1 & (1/2 \le p_1 <1)
\end{cases}
.
\ee
The lack of $T_{11}$ ($T_{22}$) for $p_1<1/2$ ($p_1 \ge 1/2$) reflects that in the maximally disassortative case, all neighbors of mask wearers (non-mask wearers) are non-mask wearers (mask wearers).
The perfectly disassortative case ($r = -1$) is realized only when $p_1=1/2$. 
In this case, the above condition becomes $T_{12}T_{21} G_q'(1)^2 =1$, so neither $T_{11}$ nor $T_{22}$ exists.

Next, we derive the probability $P_{\rm L}$ that a single seed node causes a large outbreak. 
We designate $P_n^{(x)}(n_1,n_2)$ as the probability that a seed node of type $x$ infects $n_1$ neighbors of type 1 and $n_2$ neighbors of type 2. 
This probability, $P_n^{(x)}(n_1,n_2)$, is given by 
\be
P_n^{(x)}(n_1,n_2) = \sum_{k_1=n_1}^\infty \sum_{k_2=n_2}^\infty p(k_1+k_2) \binom{k_1+k_2}{k_1} \prod_{y = 1,2} p_{xy}^{k_y} 
\binom{k_y}{n_y} T_{xy}^{n_y} (1-T_{xy})^{k_y-n_y}.
\ee
By introducing the generating function for $P_n^{(x)}(n_1,n_2)$ as $F_p^{(x)}(y,z) = \sum_{n_1} \sum_{n_2} P_n^{(x)} (n_1,n_2) y^{n_1} z^{n_2}$, we have
\begin{subequations}
\begin{align}
F_p^{(1)}(y,z) & = G_p(1+p_{11}T_{11}(y-1)+p_{12}T_{12}(z-1)), \\
F_p^{(2)}(y,z) & = G_p(1+p_{21}T_{21}(y-1)+p_{22}T_{22}(z-1)).
\end{align}
\end{subequations}
Likewise, we denote by $Q_n^{(x)}(n_1,n_2)$ the probability that a type-$x$ infected node that is reached by following a randomly chosen edge and infected through it newly infects $n_1$ type-1 neighbors and $n_2$ type-2 neighbors. 
The generating function for $Q_n^{(x)}(n_1,n_2)$ is $F_q^{(x)}(y,z)= \sum_{n_1} \sum_{n_2} Q_n^{(x)} (n_1,n_2) y^{n_1} z^{n_2}$, which is written as 
\begin{subequations}
\begin{align}
F_q^{(1)}(y,z) &= G_q(1+p_{11}T_{11}(y-1)+p_{12}T_{12}(z-1)), \\
F_q^{(2)}(y,z) &= G_q(1+p_{21}T_{21}(y-1)+p_{22}T_{22}(z-1)).
\end{align}
\end{subequations}
We further introduce the probability, $P_{s}^{(x)}(s_1,s_2)$, that a seed node of type $x$ causes an outbreak, where $s_1$ type-1 nodes and $s_2$ type-2 nodes are eventually infected, and the probability, $Q_{s}^{(x)}(s_1,s_2)$, that the infection of a type-$x$ node reached by following a randomly chosen edge results in the secondary infection of $s_1$ type-1 nodes and $s_2$ type-2 nodes. 
We define the generating function for $P_{s}^{(x)}(s_1,s_2)$ and $Q_{s}^{(x)}(s_1,s_2)$ as $H_{p}^{(x)}(y,z) = \sum_{s_1} \sum_{s_2} P_s^{(x)} (s_1,s_2) y^{s_1} z^{s_2}$ and $H_{q}^{(x)}(y,z) = \sum_{s_1} \sum_{s_2} Q_s^{(x)} (s_1,s_2) y^{s_1} z^{s_2}$, respectively. 
These generating functions obey the following equations:
\begin{subequations}
\begin{align}
H_{p}^{(1)}(y,z) &= y F_p^{(1)} (H_q^{(1)}(y,z), H_q^{(2)}(y,z)), 
\quad
H_{p}^{(2)}(y,z) = z F_p^{(2)} (H_q^{(1)}(y,z), H_q^{(2)}(y,z)), 
\\
H_{q}^{(1)}(y,z) &= y F_q^{(1)} (H_q^{(1)}(y,z), H_q^{(2)}(y,z)), 
\quad
H_{q}^{(2)}(y,z) = z F_q^{(2)} (H_q^{(1)}(y,z), H_q^{(2)}(y,z)). 
\end{align}
\end{subequations}
A randomly chosen node is of type $x$ with probability $p_x$, and it causes an epidemic with finite infections with probability $\sum_{s_1} \sum_{s_2} P_s^{(x)}(s_1,s_2)=H_p^{(x)}(1,1)$. 
Thus, the occurrence probability of a large outbreak $P_{\rm L}$, which is the probability that an epidemic starting from a single seed node does not terminate with finite infections, is 
\be
P_{\rm L}
= 1- p_1 H_p^{(1)}(1,1) - p_2 H_p^{(2)}(1,1) 
= 1- p_1 F_p^{(1)}(v_1,v_2) - p_2 F_p^{(2)}(v_1,v_2), \label{eq:pG}
\ee
where $v_1= H_q^{(1)}(1,1)$ and $v_2 = H_q^{(2)}(1,1)$ are the solutions of the following equations: 
\be
v_1 = F_q^{(1)}(v_1,v_2), 
\quad 
v_2 = F_q^{(2)}(v_1,v_2).
\ee

It should be stated that the critical transmissibility set is related to the reproductive number $R_0$, which is the average number of infections caused by a newly infected individual.
The critical transmissibility set is given as $R_0=1$ in that an epidemic persists for $R_0>1$ and dies out for $R_0<1$.
Following \cite{tian2021analysis,tian2021role}, $R_0$ is defined as as the spectral radius of matrix $B = (B_{xy})$, 
\be
R_0 = \rho(B),
\ee
where $B_{xy}$ is the expected number of type-$y$ nodes infected by a newly infected type-$x$ node.
Since the expected number of neighbors of a newly infected node is $G_q'(1)$ and each neighbor is of type $y$ with probability $p_{xy}$ and will be infected with probability $T_{xy}$, 
we have
\be
B_{xy} = G_q'(1) T_{xy} p_{xy}. \label{eq:mat}
\ee
Substituting Eq.~(\ref{eq:mat}) into $R_0 = \rho(B) = 1$, we obtain the critical transmissibility set (\ref{eq:thresholdCondition}).

Finally, we determine the average epidemic size $s_{\rm av}$ using $s_{\rm L}$ and $P_{\rm L}$. 
An infection starting from a single seed node evolves into a large epidemic outbreak of expected size $s_{\rm L}$ with probability $P_{\rm L}$; otherwise, the epidemic propagation becomes small so that the final fraction of recovered nodes is negligible. 
Thus, the average epidemic size $s_{\rm av}$ is 
\be
s_{\rm av} = P_{\rm L} s_{\rm L}
=(1- p_1 F_p^{(1)}(v_1,v_2) - p_2 F_p^{(2)}(v_1,v_2))(1- p_1 G_p(u_1) - p_2 G_p(u_2)). \label{eq:sAverage}
\ee

\section{Results \label{sec:results}}

To discuss the effect of assortative mixing by mask-wearing, we consider the mask model in two typical random networks with tunable assortative mixing: a Poisson network and a scale-free network. 
With reference to an experiment~\cite{ueki2020effectiveness}, which measured the effectiveness of mask use in filtering expelled and inhaled droplets, the present study employs the following parameters for $T_{xy}$,
\be
\begin{pmatrix}
T_{11} & T_{12} \\
T_{21} & T_{22} \\
\end{pmatrix}
=
T
\begin{pmatrix}
0.24 & 0.42 \\
0.50 & 1.00 \\
\end{pmatrix}. \label{eq:maskEffect}
\ee
We present analytical results as well as Monte Carlo (MC) simulation results. 
In the simulations, we generated $10^2$ network realizations of $n = 10^5$ nodes, and implemented the mask model with random placements of a single seed and non-mask wearers $10^3$ times on each realization. 
In each run, the size of an epidemic was given by the fraction of recovered nodes in the final state. 
Afterward, the average epidemic size $s_{\rm av}$ was given as the epidemic size averaged over all samples.
In the present simulation, we state that a large outbreak occurs if the epidemic size exceeds 0.01. 
We regard the frequency of samples in which a large outbreak occurred as the occurrence probability of large outbreaks $P_{\rm L}$, and the epidemic size averaged over those samples in which a large outbreak occurred as the mean size of large outbreaks $s_{\rm L}$. 

%%%%%%%%%%%%%%%%%%%%%%%%%%%%%%%%%%%%%
 \begin{figure}[tb]
 \centering
 (a)
  \includegraphics[width=4.75cm]{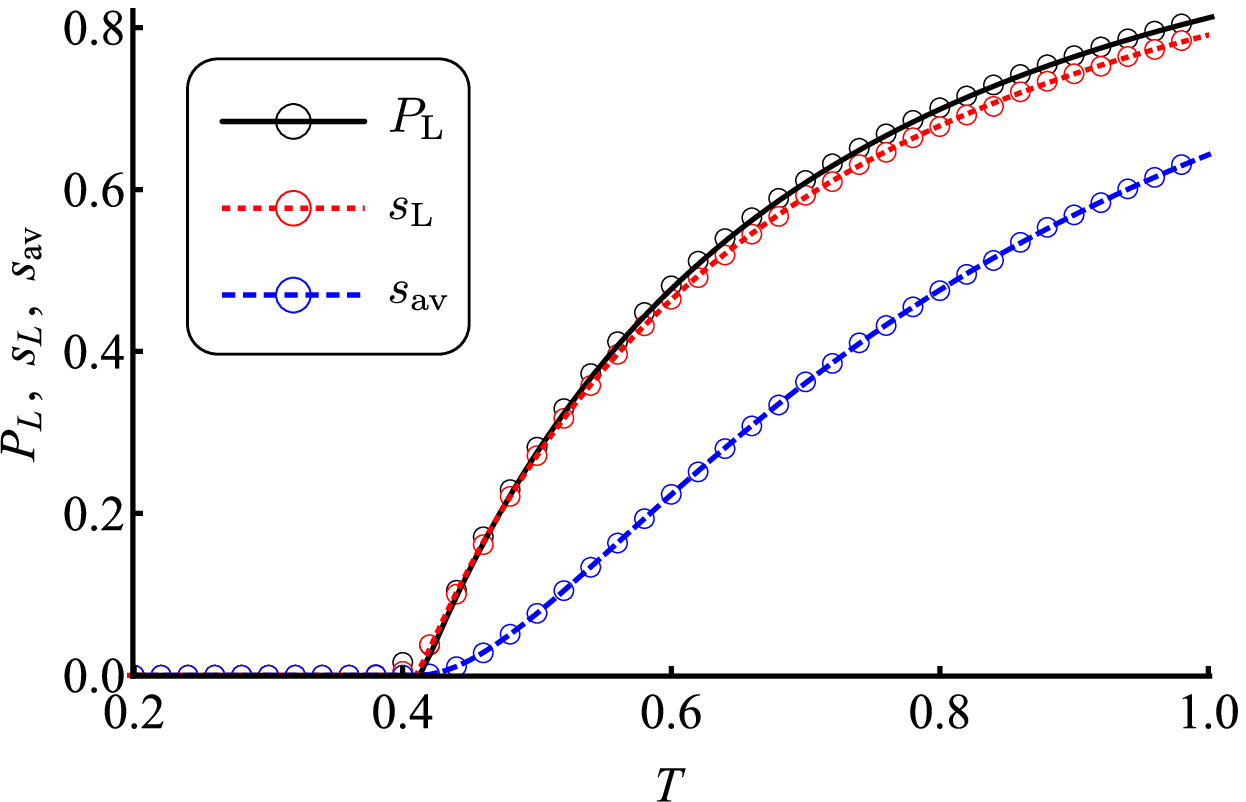}
 (b)
  \includegraphics[width=4.75cm]{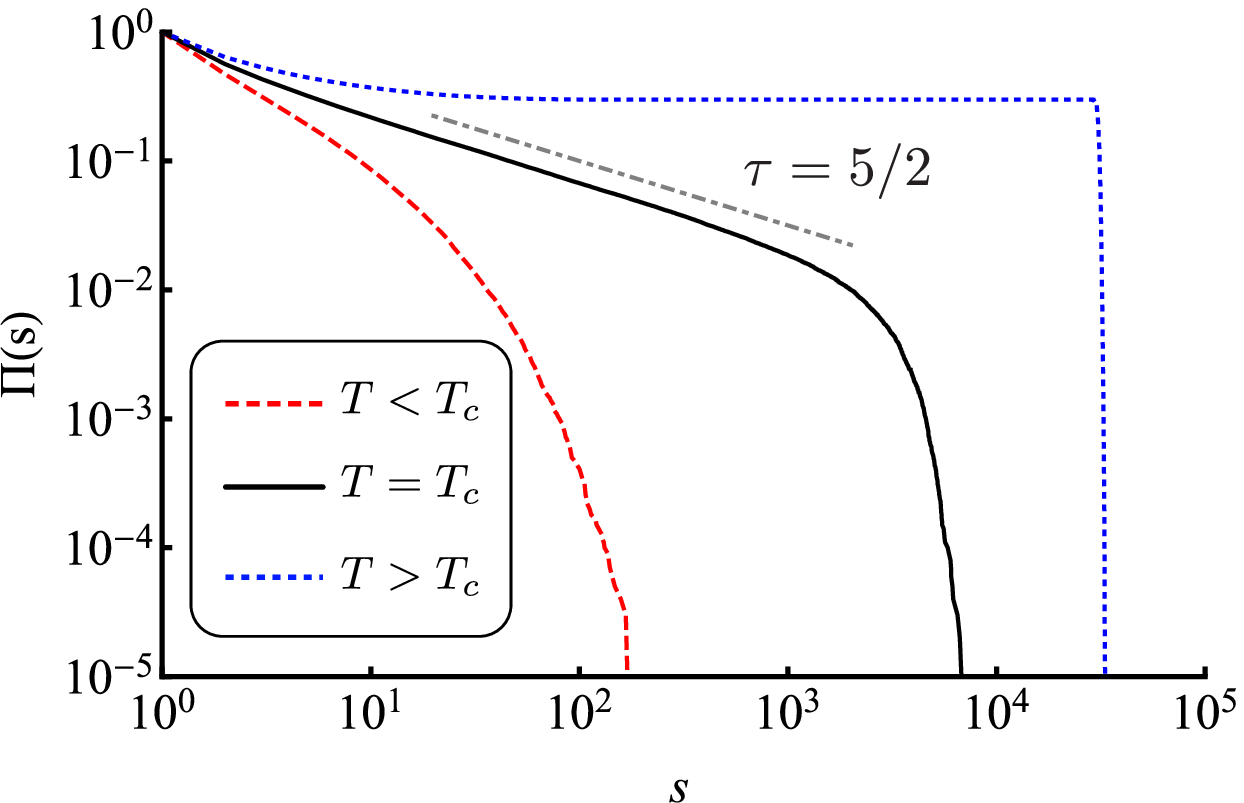}
 (c)
  \includegraphics[width=4.75cm]{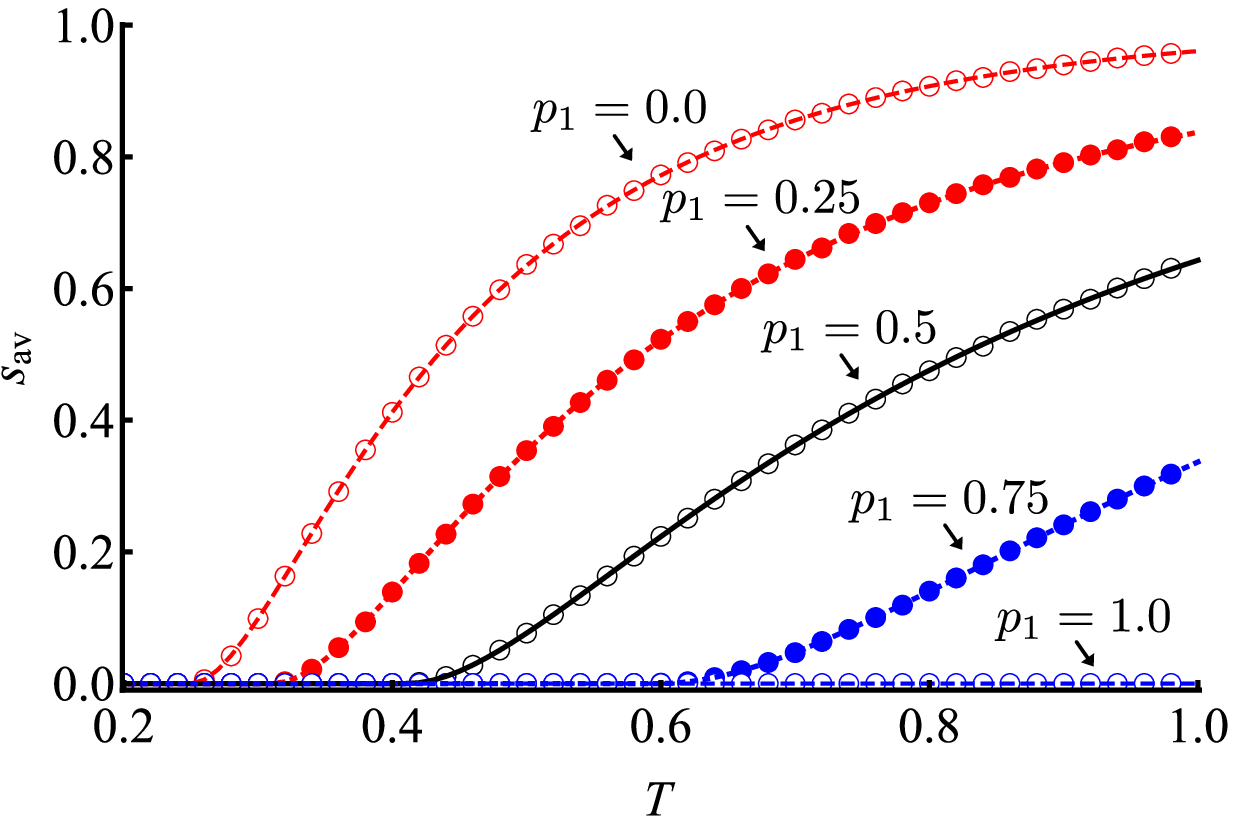}
 \caption{
Results of the mask model on the Poisson network with no assortative mixing ($r=0$). 
(a) Occurrence probability of large outbreaks $P_{\rm L}$ (black), mean size of large outbreaks $s_{\rm L}$ (red), and average epidemic size $s_{\rm av}$ (blue). 
In panel (a), the mask coverage is fixed at $p_1=0.5$. 
The symbols represent the results of MC simulations of $n=10^5$ nodes, and the lines represent the analytical results obtained from Eq.~(\ref{eq:pG}),~(\ref{eq:S}), and~(\ref{eq:sAverage}) for $P_{\rm L}$, $s_{\rm L}$, and $s_{\rm av}$, respectively. 
The analytical lines are in excellent agreement with the simulation results.
(b) The cumulative epidemic size distribution, $\Pi(s)=\sum_{s'=s}^\infty \pi(s')$, obtained from MC simulations with $p_1=0.5$ and with $T=0.3114<T_c$ (red-dashed line), $T=0.4114 \approx T_c$ (black-solid line), and $T=0.5114>T_c$ (blue-dotted line). 
Here $\pi(s)$ denotes the ratio of the samples whose epidemic size is $s$ in $10^2 \times 10^3$ samples. 
Panel (b) indicates that $\Pi(s)$ quickly decreases with $s$ for $T<T_c$, $\Pi(s) \propto s^{2-\tau}$ at $T=T_c$, and $\Pi(s)$ shows the bimodality of $\pi(s)$ for $T>T_c$.
(c) Impact of the mask coverage ($p_1=0.0,\, 0.25,\, 0.5,\, 0.75$, and $1.0$) on the average epidemic size $s_{\rm av}$. 
The symbols represent the simulation results, and the lines represent the analytical results based on Eq.~(\ref{eq:sAverage}). 
The increase in mask coverage decreases the average epidemic size and increases the epidemic threshold.
 }
 \label{fig:Examples}
 \end{figure}
%%%%%%%%%%%%%%%%%%%%%%%%%%%%%%%%%%%%%

First, we treat the mask model on the Poisson network whose degree distribution is $p(k)=\langle k \rangle^k e^{-\langle k \rangle}/k!$ and average degree is $\langle k \rangle=4$. 
Substituting Eq.~(\ref{eq:maskEffect}) and $G_q'(1)=\langle k^2 - k \rangle/\langle k \rangle=\langle k \rangle$ into Eq.~(\ref{eq:thresholdCondition}), it follows that the epidemic threshold $T_c$ is given as the solution of $(0.96(r+p_1(1-r))T_c-1)(4(1-p_1(1-r))T_c-1)-3.36 p_1(1-p_1)(1-r)^2 T_c^2=0$.
We begin by considering the case of no mixing ($r = 0$).
Figure~\ref{fig:Examples} (a) shows the average epidemic size $s_{\rm av}$, occurrence probability of large outbreaks $P_{\rm L}$, and mean size of large outbreaks $s_{\rm L}$ when the mask coverage is fixed at $p_1=0.5$. 
The epidemic threshold is $T_c \approx 0.4114$ when $p_1=0.5$ and $r=0$.
Evidently, all the theoretical lines drawn from equations in the previous section are in excellent agreement with the simulation data (symbols).

As shown in Fig.~\ref{fig:Examples} (a), all the $s_{\rm av}$, $s_{\rm L}$, and $P_{\rm L}$ parameters exhibit non-zero values when $T$  is greater than the epidemic threshold $T_c $.
Further, we observe from the distribution of the epidemic sizes that a phase transition occurs at $T=T_c$. 
Figure~\ref{fig:Examples} (b) shows the cumulative distribution of epidemic sizes, $\Pi(s)=\sum_{s'=s}^\infty \pi(s')$, where $\pi(s)$ is the frequency of samples in which the epidemic size is $s$, when $T =0.3114 \;(< T_c)$, $T=0.4114 \;(\approx T_c)$, and $T=0.5114 \;(>T_c)$. 
The figure indicates that $\Pi(s)$ for $T<T_c$ quickly decreases with $s$; 
$\Pi(s)$ for $T>T_c$ stands for the bimodality of $\pi(s)$ with peaks at small $s$ ($s < 10^{-3}$) and at large $s$ ($s > 10^{-1}$) values; 
and $\Pi(s)$ for $T=T_c$ follows the power law, $\Pi(s) \propto s^{2-\tau}$. 
In the present case, exponent $\tau$ is given as $\tau=5/2$, which coincides with that in the mean-field theory of percolation \cite{stauffer2018introduction} (and the SIR model is known to be in percolation universality class \cite{grassberger1983critical,tome2010critical,de2011new}). 

Figure~\ref{fig:Examples} (c) indicates the impact of the mask coverage in suppressing the average epidemic size $s_{\rm av}$. 
As expected, the increase in the number of mask wearers decreases the average epidemic size and also increases the epidemic threshold. 
Furthermore, a high mask coverage results in lowered $P_{\rm L}$ and $s_{\rm L}$ (not shown), i.e., the higher the number of mask wearers, the lower the occurrence probability of a large outbreak, and even if it does occur, the mask wearers will reduce its propagation.

%%%%%%%%%%%%%%%%%%%%%%%%%%%%%%%%%%%%%
 \begin{figure}[tb]
 \centering
(a)
  \includegraphics[width=6.5cm]{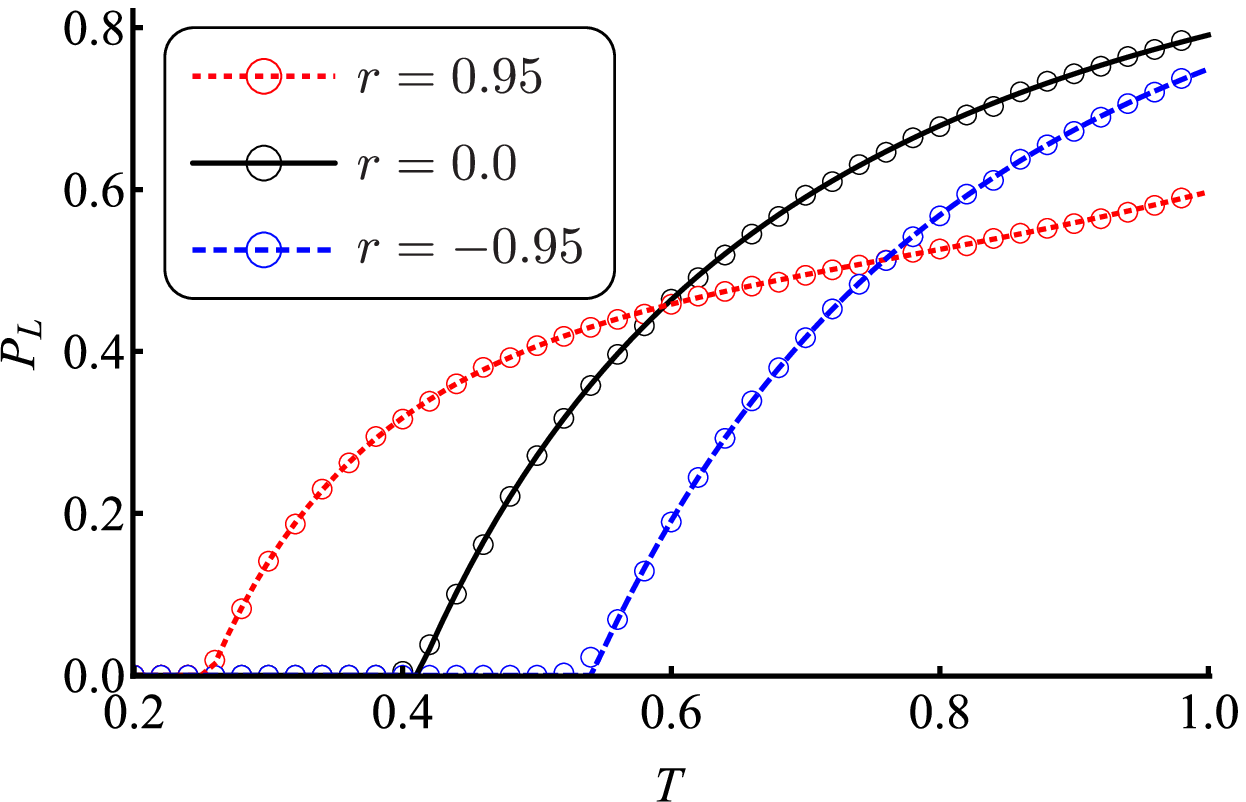}
(b)
  \includegraphics[width=6.5cm]{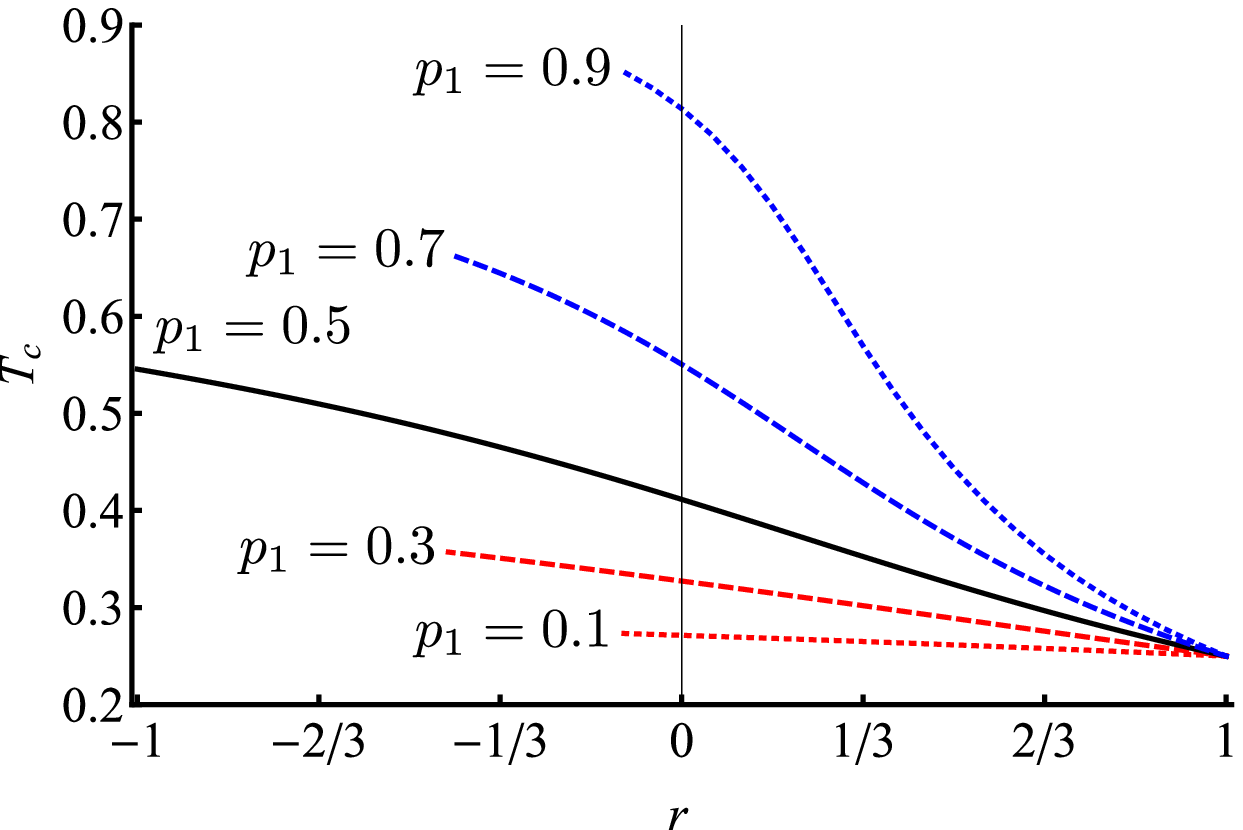}
 \caption{
(a) Occurrence probability $P_{\rm L}$ of the mask model in the Poisson network with different assortative mixing patterns: assortative ($r=0.95$; red), neutral ($r=0$; black), and disassortative cases ($r=-0.95$; blue). 
The symbols represent the simulation results of $n = 10^5$ nodes, and the lines represent the analytical results based on Eq.~(\ref{eq:pG}). 
The epidemic threshold above which $P_{\rm L}$ becomes nonzero is the lowest (highest) in the assortative (disassortative) case.
On the other hand, $P_{\rm L}$ for high-transmissibility cases ($T \approx 1$) is smaller in the assortative case than in the disassortative case (the situation is the same for $s_{\rm L}$ and $s_{\rm av}$). 
Assortative mixing reduces the probability and degree of a large outbreak in high-transmissibility cases, although it allows a large outbreak to occur even in low-transmissibility cases. 
(b) Epidemic threshold $T_c$ as a function of $r$ when $p_1=0.1,\, 0.3,\, 0.5,\, 0.7$, and $0.9$. 
Here $T_c$ is evaluated from Eq.~(\ref{eq:thresholdCondition}). 
It is obvious that $T_c$ continuously decreases as $r$ increases.
 }
 \label{fig:severalMaskPatterns}
 \end{figure}
%%%%%%%%%%%%%%%%%%%%%%%%%%%%%%%%%%%%%

Next, we discuss the impact of the assortative mixing by mask-wearing on epidemic spreadings, fixing the mask coverage at $p_1=0.5$ (where the assortativity coefficient $r$ is in the range of $[-1,1]$). 
Figure~\ref{fig:severalMaskPatterns} (a) shows a comparison of the occurrence probabilities for the assortative ($r=0.95$), neutral ($r=0$), and disassortative cases ($r=-0.95$). 
As indicated by both the theoretical line and simulation data, the epidemic threshold is the lowest (highest) when the mixing based on mask-wearing is assortative (disassortative).
Disassortative mixing is effective in suppressing epidemics in that a high transmissibility is required for large outbreaks.  
On the other hand, $P_{\rm L}$ for high-transmissibility cases ($T \approx 1$) is smaller in the assortative case than in the disassortative case (the situation is the same for $s_{\rm L}$ and $s_{\rm av}$). 
In the assortative case where non-mask wearers tend to be connected, large outbreaks are more likely to occur in the groups of non-mask wearers. 
Simultaneously, in the assortative case, mask wearers are often surrounded by mask wearers, and there is a good probability of containing the epidemic in the early stages, if the seed node is a mask wearer, even with high transmissibility. 
The separation of mask wearers and non-mask wearers reduces the probability and degree of a large outbreak in high-transmissibility cases, although it allows a large outbreak to occur even in low-transmissibility cases. 
Note that the value of $P_{\rm L}$ (also $s_{\rm L}$ and $s_{\rm av}$) in a high $T$ case is affected by the mask effectiveness, in addition to assortative mixing by mask-wearing, and mask coverage.

%%%%%%%%%%%%%%%%%%%%%%%%%%%%%%%%%%%%%
 \begin{figure}[tb]
 \centering
  (a)
  \includegraphics[height=6.5cm]{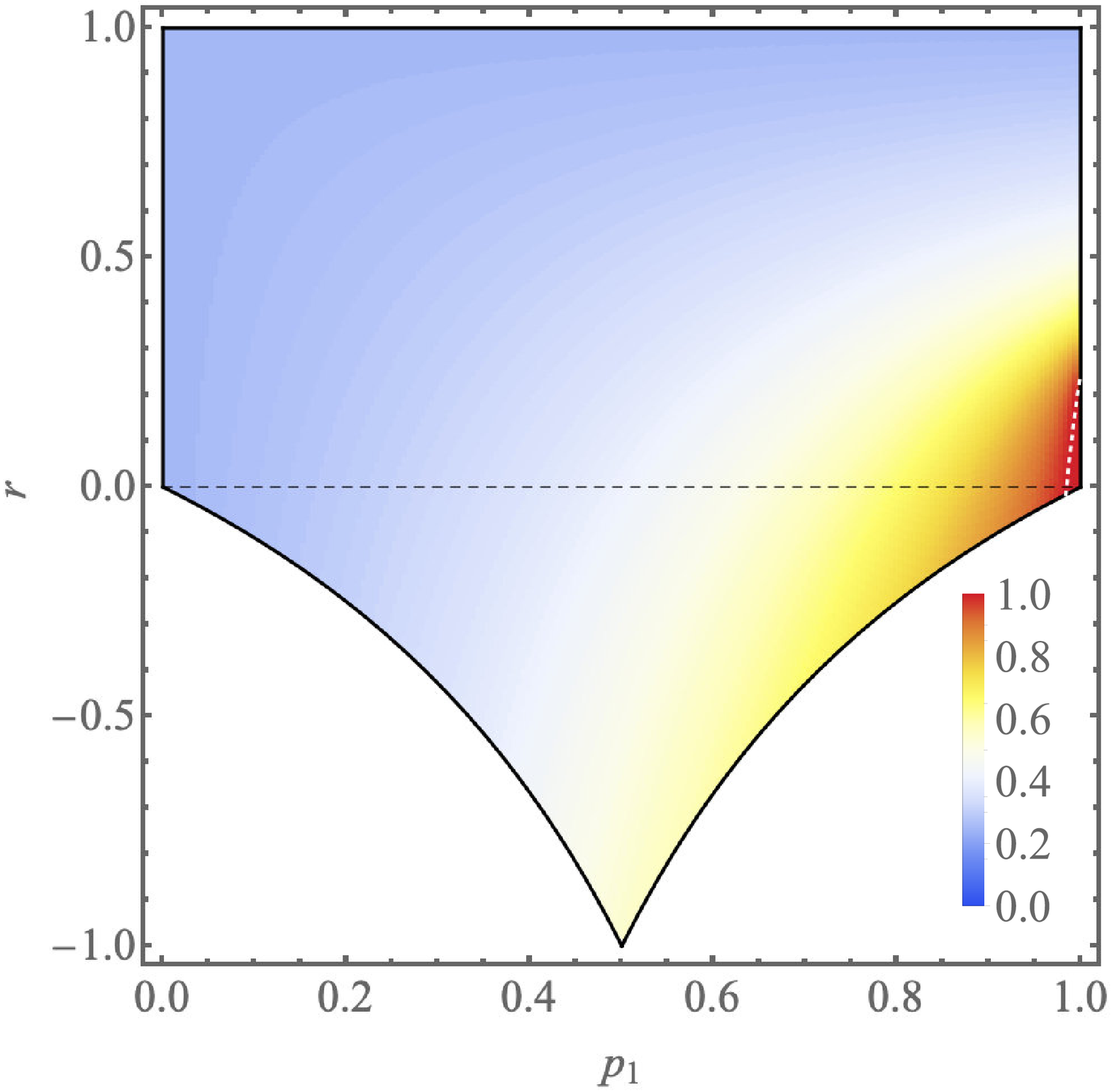}
  (b)
  \includegraphics[height=6.5cm]{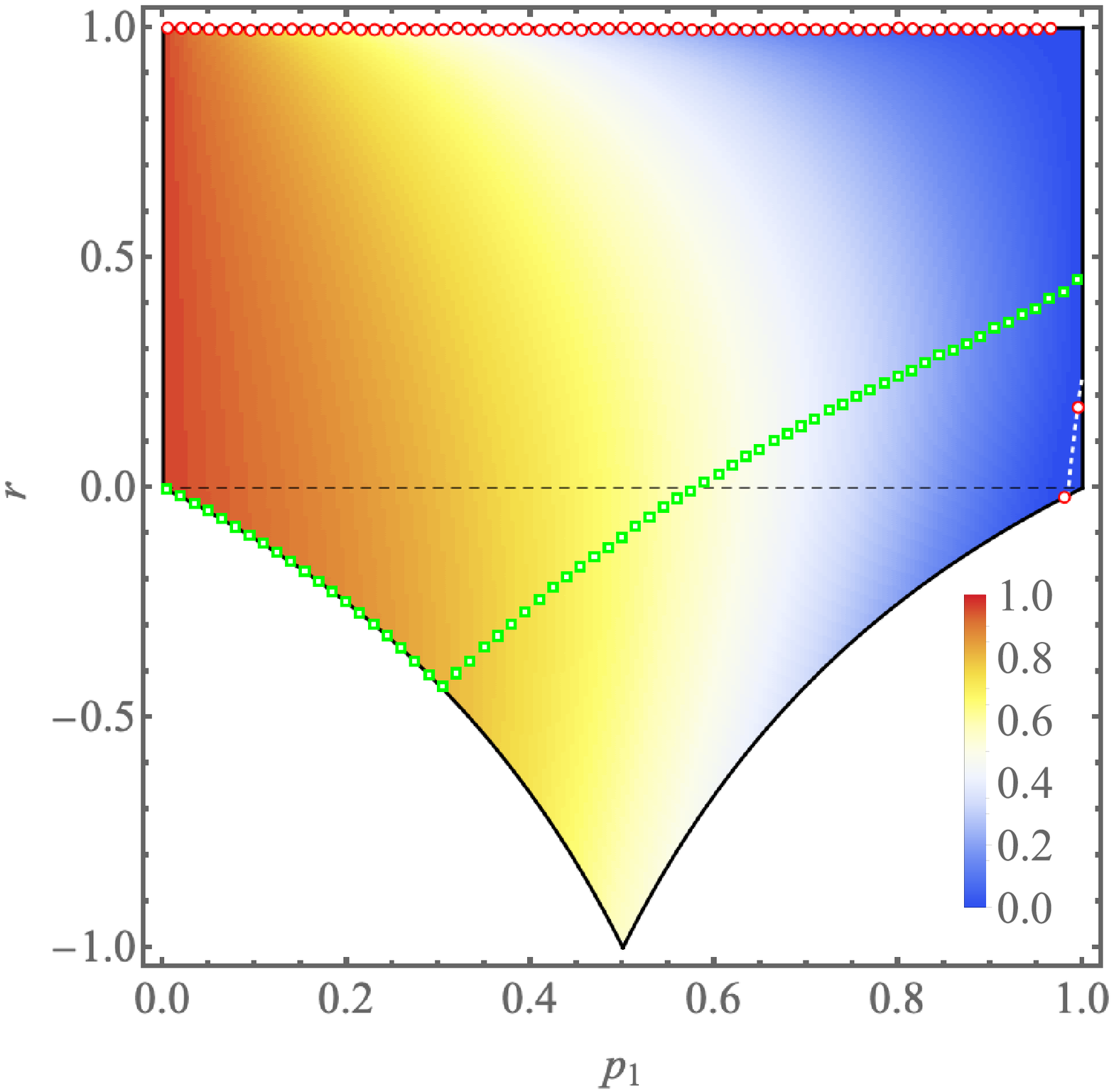}
 \caption{
Impact of assortative mixing on (a) epidemic threshold $T_c$ and (b) average epidemic size $s_{\rm av}$ with $T=1$ in the $(p_1, r)$ plane. 
Here, the color-coded values of $T_c$ and $s_{\rm av}$ are evaluated from Eq.~(\ref{eq:thresholdCondition}) and Eq.~(\ref{eq:sAverage}), respectively. 
The black curved lines represent $r_{\rm min}(p_1)$ given as Eq.~(\ref{eq:rmin}).
In panel (a), $T_c$ has a minimum (maximum) value at $r=1$ ($r=r_{\rm min}$) for a given mask coverage $p_1$ ($0<p_1<1$).
Panel (a) illustrates that $T_c$ is generally increased by decreasing $r$ or by increasing $p_1$.
The red circles (the green squares) in panel (b) indicate the position of $r$ at which $s_{\rm av}$ reaches the minimum (maximum) for a given $p_1$. 
On the right side of the white dashed line, a large outbreak does not occur even when $T=1$ (i.e., $T_c=1$ and $s_{\rm av}=0$).
Irrespective of $p_1$, $s_{\rm av}$ reaches the minimum in the strongly assortative case ($r \approx 1$).
On the other hand, $s_{\rm av}$ reaches the maximum in a moderately assortative case when the mask coverage is high ($p_1 \gtrsim 0.55$).
}
 \label{fig:colorMap}
 \end{figure}
%%%%%%%%%%%%%%%%%%%%%%%%%%%%%%%%%%%%%

Assortative mixing by mask-wearing alters the epidemic threshold. 
Figure~\ref{fig:severalMaskPatterns} (b) plots the epidemic threshold $T_c$ as a function of the assortativity coefficient $r$ with different mask coverages: $p_1 = 0.1,\, 0.3,\, 0.5,\, 0.7$, and $0.9$. 
It is obvious that the epidemic threshold continuously decreases with increasing $r$. 
As $r$ increases, the separation of mask wearers and non-mask wearers becomes more pronounced, resulting in large outbreaks among non-mask wearers more easily.
At $r = 1$, regardless of the value of $p_1(<1)$, non-mask wearers are completely detached from mask wearers and form their own giant component~\footnote{In assortatively mixed random networks, even if the fraction of non-mask wearers is small, as long as $p_1<1$, the number of non-mask wearers is infinite and the degree distribution of non-mask wearers' component follows $p(k)$, thus a giant component can be formed.}, so the epidemic threshold becomes determined by whether the infection spreads among non-mask wearers, which is consistent with the known one (\ref{eq:knownThreshold}) (see also Eq.~(\ref{eq:thresholdAssortative})).
In addition, we find that the impact of assortative mixing on the epidemic threshold increases as the mask coverage increases. 
Figure~\ref{fig:colorMap} (a) illustrates the color-coded values of $T_c$ in the $(p_1,r)$ plane. 
The figure shows that the epidemic threshold is generally increased by disassortative mixing by mask-wearing or by increasing the mask coverage.
For each $p_1$ value ($0<p_1<1$), $T_c$ has a minimum (maximum) value at $r=1$ ($r=r_{\rm min}$).

Although the assortative mixing lowers the epidemic threshold, it suppresses $P_{\rm L}$, $s_{\rm L}$, and $s_{\rm av}$ when the transmissibility is high, as already mentioned in Fig.~\ref{fig:severalMaskPatterns} (a). 
Figure~\ref{fig:colorMap} (b) shows the average epidemic size $s_{\rm av}$ at $T=1$.
Irrespective of $p_1$, $s_{\rm av}$ reaches the minimum in the strongly assortative case ($r \approx 1$).
That is, the mask use is most effective in terms of reducing $P_{\rm L}$ when $r \approx 1$.
 This, however, does not imply that the average epidemic size for high $T$ always decreases more when the mixing is assortative than when it is neutral or disassortative.
When the mask coverage is high ($p_1 \gtrsim 0.55$), $s_{\rm av}$ reaches the maximum in a moderately assortative case (see the green line in Fig.~\ref{fig:colorMap} (b)).
The same holds for $P_{\rm L}$ and $s_{\rm L}$ (not shown). 

\begin{figure}[tb]
 \centering
 \vspace{3cm}
(a) 
 \includegraphics[width=4.75cm]{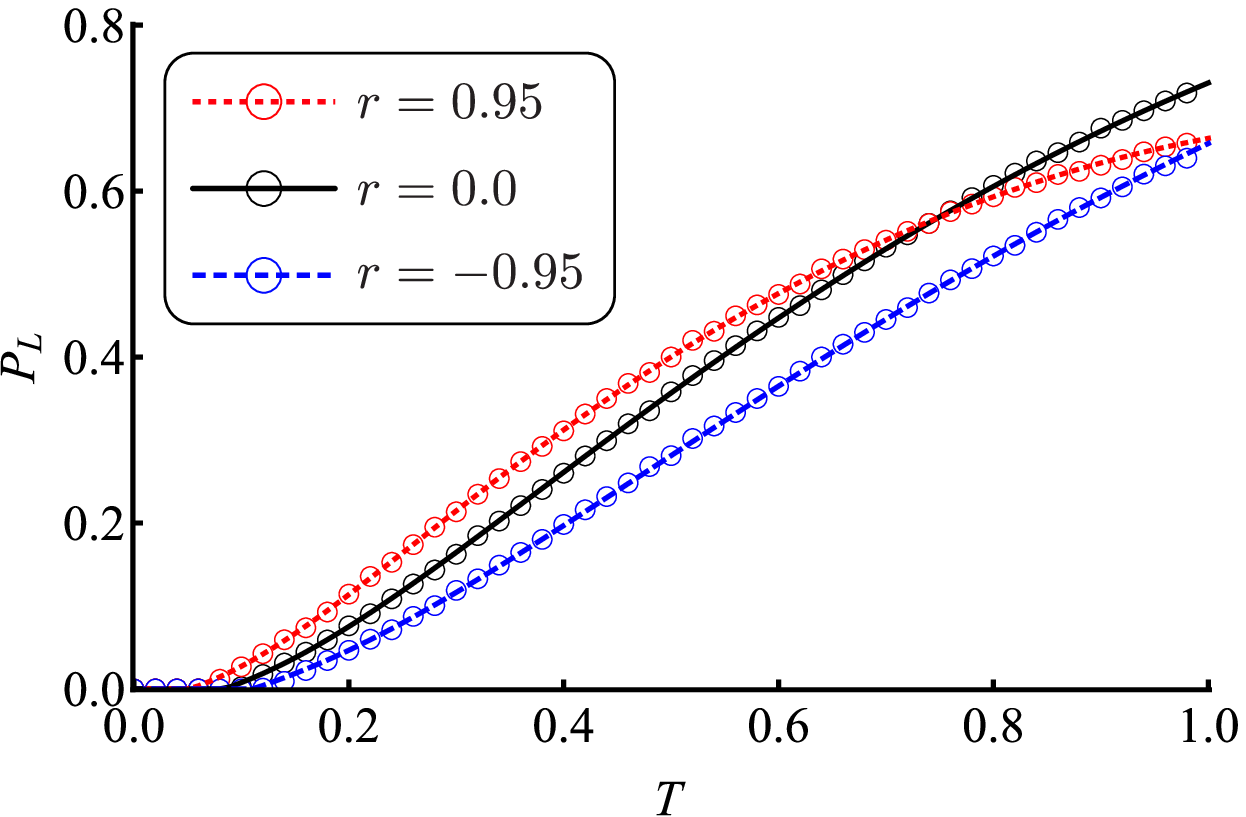}
(b)
 \includegraphics[width=4.75cm]{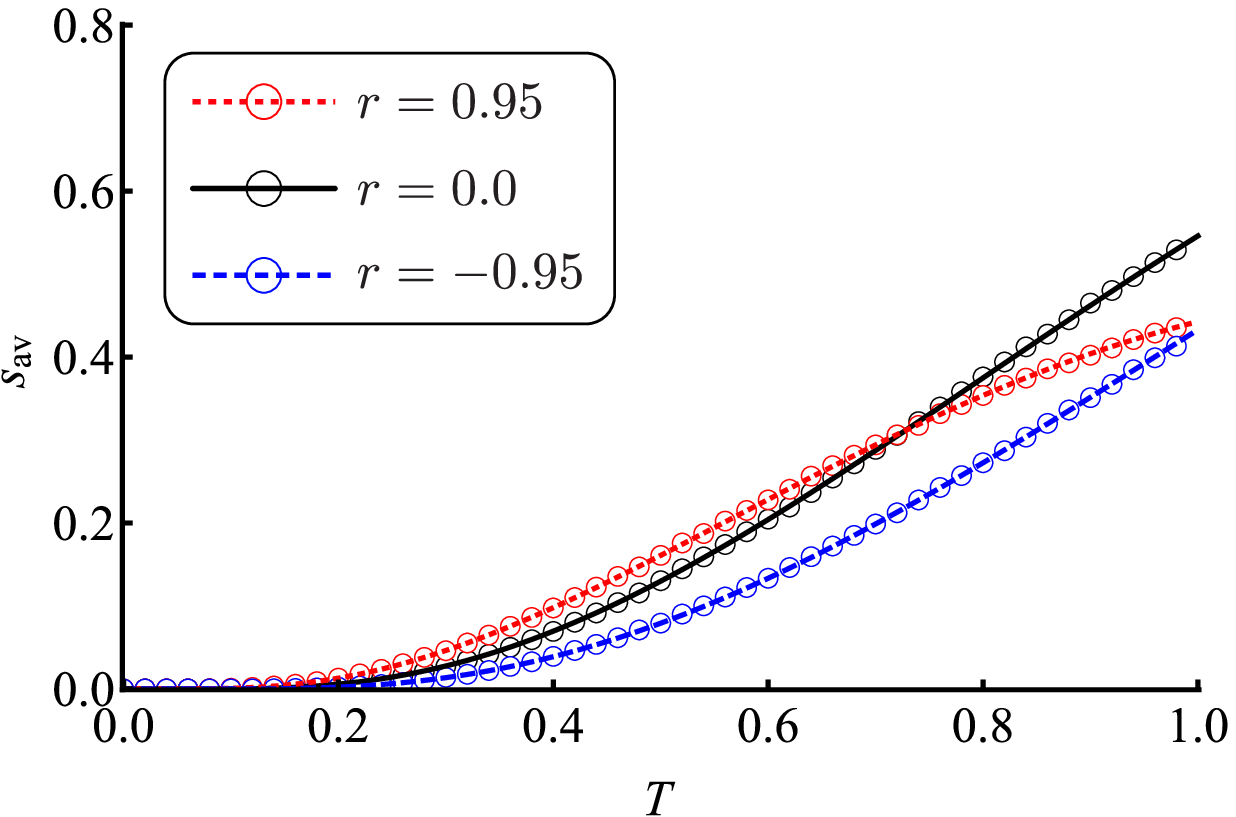}
(c)
\includegraphics[width=4.0cm]{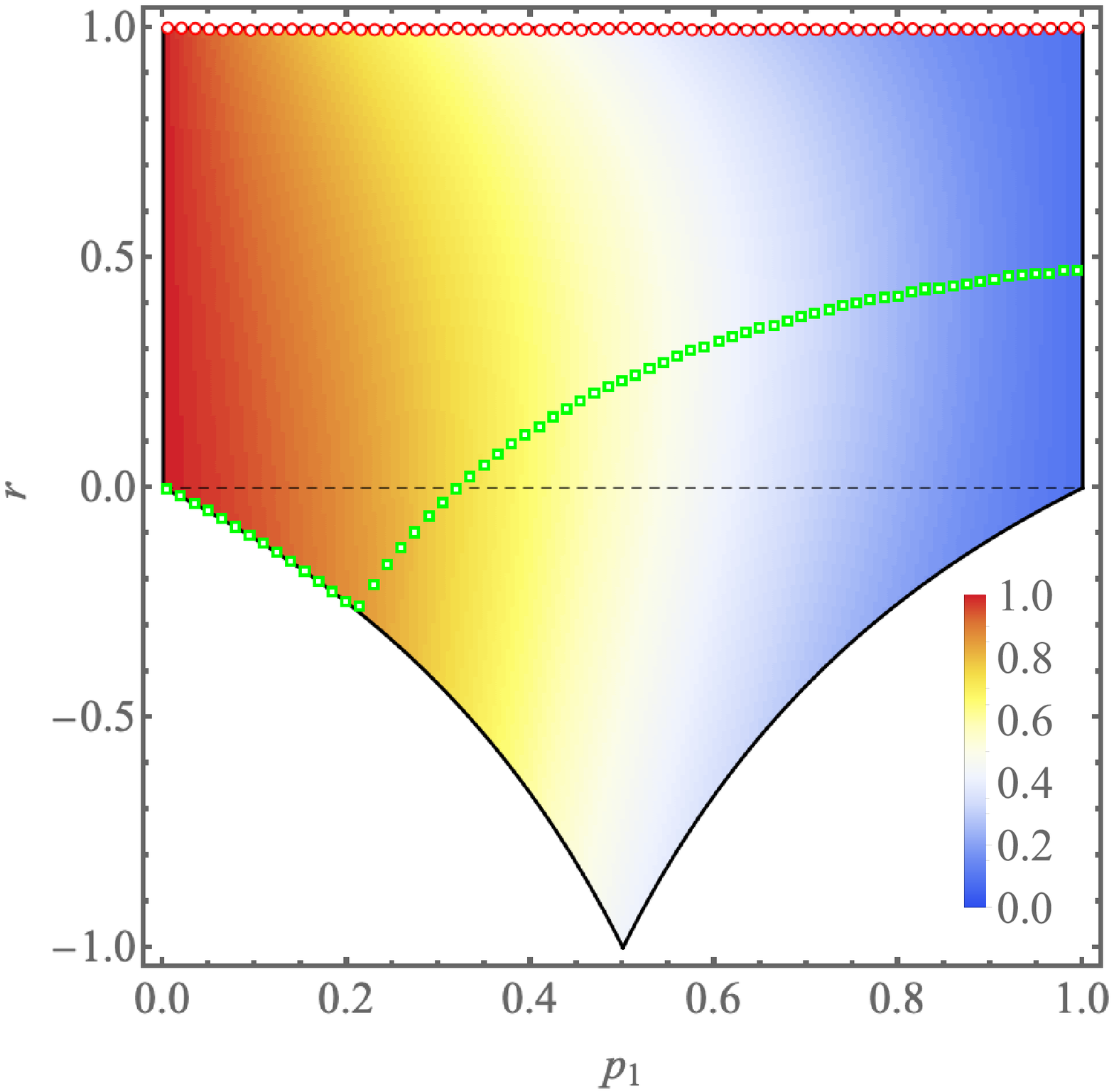}
 \caption{
Results of the mask model in a scale-free network: 
(a) occurrence probability of large outbreaks $P_{\rm L}$ and (b) average epidemic size $s_{\rm av}$ as a function of $T$ in the assortative ($r=0.95$; red), neutral ($r=0$; black), and disassortative cases ($r=-0.95$; blue). 
The symbols represent the simulation results of $n = 10^5$ nodes, and the lines represent the analytical results based on Eq.~(\ref{eq:pG}) for panel (a) and Eq.~(\ref{eq:sAverage}) for panel (b). 
The degree distribution is $p(k) = k^{-2.625} / \sum_{k'=k_{\rm min}}^{k_{\rm max}} k'^{-2.625}$ for $k_{\rm min} \le k \le k_{\rm max}$, where $k_{\rm min}=2$ and $k_{\rm max}=500$.
The epidemic threshold is the lowest (highest) in the assortative (disassortative) case, although it is comparatively low regardless of the mixing pattern.
The occurrence probability and the average epidemic size in high $T$ cases are almost identical in the assortative and disassortative cases. 
(c) Color-coded values of $s_{\rm av}$ at $T=1$ in the $(p_1,\;r)$ plane.
The black curved lines in panel (c) represent $r_{\rm min}(p_1)$, which is given as Eq.~(\ref{eq:rmin}).
The red circles (the green squares) in panel (c) indicate the position of $r$ at which $s_{\rm av}$ reaches the minimum (maximum) for a given $p_1$.
The average epidemic size at $T=1$ is minimized in the strongly assortative case ($r \approx 1$), but it also exhibits an equally small value in the maximally disassortative case ($r \approx r_{\rm min}$), provided that the mask coverage is not low.
 }
 \label{fig:heterogeneousNetwork}
 \end{figure}

Finally, we consider the example of heterogeneous networks. 
We employ a scale-free random network having $p(k) = k^{-\gamma} / \sum_{k'=k_{\rm min}}^{k_{\rm max}} k'^{-\gamma}$ ($k=k_{\rm min}, k_{\rm min}+1, \ldots, k_{\rm max}$). 
Here, we set $k_{\rm min}=2$, $k_{\rm max}=500$, and $\gamma=2.625$ so that $\langle k \rangle \approx 4$. 
Although the epidemic threshold for a scale-free network with finite $k_{\rm max}$ is comparatively low regardless of the assortative mixing pattern, it is the lowest in the assortative case (Figs.~\ref{fig:heterogeneousNetwork} (a) and (b)). 
(Note that this small but nonzero epidemic threshold is due to the finiteness of $k_{\rm max}$. If $k_{\rm max}$ is unbounded, $T_c$ becomes zero, regardless of the assortative mixing pattern, since $G_q'(1)$ diverges.)
Figures~\ref{fig:heterogeneousNetwork} (a) and (b) reveal that the occurrence probability and the average epidemic size in high $T$ cases are almost identical in the assortative and disassortative cases. 
We see from Fig.~\ref{fig:heterogeneousNetwork} (c) that the average epidemic size at $T=1$ is minimized in the strongly assortative case ($r \approx 1$), but it also exhibits an equally small value in the maximally disassortative case ($r \approx r_{\rm min}$), provided that the mask coverage is not low. 
Conversely, the mask use is less effective when the network is moderately assortative (as evidenced by the large $s_{\rm av}$), as in the case of Poisson networks.
We conclude that the effectiveness of mask use is enhanced when the mixing pattern is strongly assortative so that the network is split into two groups of mask wearers and non-mask wearers, or when the mixing pattern is strongly disassortative, which reduces the number of high-risk edges between non-mask wearers significantly.

%%%%%%%%%%%%%%%%%%%%%%%%%%%%%%%%%%%%%
%%%%%%%%%%%%%%%%%%%%%%%%%%%%%%%%%%%%%

\section{Summary \label{sec:summary}}

In this work, we studied the mask model, which is a SIR-type epidemic model involving mask wearers and non-mask wearers, in assortatively mixed random networks to discuss the impact of assortative mixing by mask-wearing on the effectiveness of mask use. 
We derived the occurrence probability and mean size of large outbreaks, epidemic threshold, and average epidemic size using the probability generating functions. 
In both the Poisson and scale-free random networks, we established that the assortative mixing by mask-wearing decreases the epidemic threshold.
It means that large outbreaks can occur more easily as a network is assortative.
In high-transmissibility cases, on the other hand, the average epidemic size is minimized when the mixing pattern is strongly assortative.  
Strongly assortative mixing, resulting in the separation of mask wearers and non-mask wearers, reduces the probability and degree of a large outbreak in high-transmissibility cases, although it allows a large outbreak to occur even in low-transmissibility cases.
In scale-free networks, the mask use in the maximally disassortative case is equally effective in terms of reducing the probability and mean size of large outbreaks as in the strongly assortative case (when the mask coverage is not low).
We also performed MC simulations of the mask model in the Poisson and scale-free networks, which validated our analytical treatment results. 

This study has the following limitations:
(i) the epidemic model employed is a discrete-time SIR model with constant transmissibility, which is very simple.
The effectiveness of the mask use should be discussed further in more general epidemic models.
(ii) We assumed random networks. 
Real networks have structural properties, such as degree correlation, high clustering, and community structure.
A recent study~\cite{duh2019assortativity} reported that degree correlation in a network affects human behaviors, which may be related to epidemic control.
A further study of how the combination of network structure and mixing pattern by mask-wearing affects the spread of the disease should be conducted.
Furthermore, we may consider multiplex or multilayer networks~\cite{boccaletti2014structure, bianconi2018multilayer} or simplicial complexes~\cite{torres2020simplicial, bianconi2021higher}.
Using recent bond percolation literatures~\cite{kryven2017finite,kryven2019bond,bianconi2019percolation,balogh2020networks}, we could investigate mask models on such networks.
(iii) The present study discussed the impact of assortative mixing by mask-wearing. 
One can consider the assortative mixing based on other attributes, such as age, gender, obesity, and vaccination~\footnote{After the submission of this manuscript, the impacts of vaccination homophily (assortative mixing by vaccination adoptions) on disease dynamics have been reported~\cite{hiraoka2021herd,burgio2022homophily}.}. 
Regarding the mask use, one may further consider the mask quality (cotton mask, surgical mask, or N-95 mask) as an attribute rather than usage~\footnote{In preparation of this manuscript, Tian et al.~\cite{tian2021role} proposed a multi-type mask model, where each node wears a different type of mask.}. 
The analysis provided in this study will be easily generalized to discuss some of the limitations mentioned above.
It is hoped that the present study will contribute to a better understanding of epidemic propagation patterns in our society.

\section*{Acknowledgement}
This work was supported by JSPS KAKENHI Grant Numbers JP19K03648 and JP21H03425.


\begin{thebibliography}{58}
\expandafter\ifx\csname natexlab\endcsname\relax\def\natexlab#1{#1}\fi
\expandafter\ifx\csname bibnamefont\endcsname\relax
  \def\bibnamefont#1{#1}\fi
\expandafter\ifx\csname bibfnamefont\endcsname\relax
  \def\bibfnamefont#1{#1}\fi
\expandafter\ifx\csname citenamefont\endcsname\relax
  \def\citenamefont#1{#1}\fi
\expandafter\ifx\csname url\endcsname\relax
  \def\url#1{\texttt{#1}}\fi
\expandafter\ifx\csname urlprefix\endcsname\relax\def\urlprefix{URL }\fi
\providecommand{\bibinfo}[2]{#2}
\providecommand{\eprint}[2][]{\url{#2}}

\bibitem[{\citenamefont{Estrada}(2020)}]{estrada2020covid}
\bibinfo{author}{\bibfnamefont{E.}~\bibnamefont{Estrada}},
  \bibinfo{journal}{Physics Reports}  (\bibinfo{year}{2020}).

\bibitem[{\citenamefont{Barab{\'a}si}(2016)}]{barabasi2016network}
\bibinfo{author}{\bibfnamefont{A.-L.} \bibnamefont{Barab{\'a}si}},
  \emph{\bibinfo{title}{Network Science}} (\bibinfo{publisher}{Cambridge
  University Press}, \bibinfo{year}{2016}).

\bibitem[{\citenamefont{Newman}(2018)}]{newman2018networks}
\bibinfo{author}{\bibfnamefont{M.}~\bibnamefont{Newman}},
  \emph{\bibinfo{title}{Networks}} (\bibinfo{publisher}{Oxford University
  Press}, \bibinfo{year}{2018}).

\bibitem[{\citenamefont{Pastor-Satorras
  et~al.}(2015)\citenamefont{Pastor-Satorras, Castellano, Van~Mieghem, and
  Vespignani}}]{pastor2015epidemic}
\bibinfo{author}{\bibfnamefont{R.}~\bibnamefont{Pastor-Satorras}},
  \bibinfo{author}{\bibfnamefont{C.}~\bibnamefont{Castellano}},
  \bibinfo{author}{\bibfnamefont{P.}~\bibnamefont{Van~Mieghem}},
  \bibnamefont{and}
  \bibinfo{author}{\bibfnamefont{A.}~\bibnamefont{Vespignani}},
  \bibinfo{journal}{Reviews of Modern Physics} \textbf{\bibinfo{volume}{87}},
  \bibinfo{pages}{925} (\bibinfo{year}{2015}).

\bibitem[{\citenamefont{Pastor-Satorras and
  Vespignani}(2001)}]{pastor2001epidemic}
\bibinfo{author}{\bibfnamefont{R.}~\bibnamefont{Pastor-Satorras}}
  \bibnamefont{and}
  \bibinfo{author}{\bibfnamefont{A.}~\bibnamefont{Vespignani}},
  \bibinfo{journal}{Physical Review Letters} \textbf{\bibinfo{volume}{86}},
  \bibinfo{pages}{3200} (\bibinfo{year}{2001}).

\bibitem[{\citenamefont{Moreno et~al.}(2002)\citenamefont{Moreno,
  Pastor-Satorras, and Vespignani}}]{moreno2002epidemic}
\bibinfo{author}{\bibfnamefont{Y.}~\bibnamefont{Moreno}},
  \bibinfo{author}{\bibfnamefont{R.}~\bibnamefont{Pastor-Satorras}},
  \bibnamefont{and}
  \bibinfo{author}{\bibfnamefont{A.}~\bibnamefont{Vespignani}},
  \bibinfo{journal}{The European Physical Journal B-Condensed Matter and
  Complex Systems} \textbf{\bibinfo{volume}{26}}, \bibinfo{pages}{521}
  (\bibinfo{year}{2002}).

\bibitem[{\citenamefont{Markovi{\v{c}}
  et~al.}(2021)\citenamefont{Markovi{\v{c}}, {\v{S}}terk, Marhl, Perc, and
  Gosak}}]{markovivc2021socio}
\bibinfo{author}{\bibfnamefont{R.}~\bibnamefont{Markovi{\v{c}}}},
  \bibinfo{author}{\bibfnamefont{M.}~\bibnamefont{{\v{S}}terk}},
  \bibinfo{author}{\bibfnamefont{M.}~\bibnamefont{Marhl}},
  \bibinfo{author}{\bibfnamefont{M.}~\bibnamefont{Perc}}, \bibnamefont{and}
  \bibinfo{author}{\bibfnamefont{M.}~\bibnamefont{Gosak}},
  \bibinfo{journal}{Results in Physics} \textbf{\bibinfo{volume}{26}},
  \bibinfo{pages}{104433} (\bibinfo{year}{2021}).

\bibitem[{\citenamefont{Gallos et~al.}(2007)\citenamefont{Gallos, Liljeros,
  Argyrakis, Bunde, and Havlin}}]{gallos2007improving}
\bibinfo{author}{\bibfnamefont{L.~K.} \bibnamefont{Gallos}},
  \bibinfo{author}{\bibfnamefont{F.}~\bibnamefont{Liljeros}},
  \bibinfo{author}{\bibfnamefont{P.}~\bibnamefont{Argyrakis}},
  \bibinfo{author}{\bibfnamefont{A.}~\bibnamefont{Bunde}}, \bibnamefont{and}
  \bibinfo{author}{\bibfnamefont{S.}~\bibnamefont{Havlin}},
  \bibinfo{journal}{Physical Review E} \textbf{\bibinfo{volume}{75}},
  \bibinfo{pages}{045104} (\bibinfo{year}{2007}).

\bibitem[{\citenamefont{Wang et~al.}(2016)\citenamefont{Wang, Bauch,
  Bhattacharyya, d'Onofrio, Manfredi, Perc, Perra, Salath{\'e}, and
  Zhao}}]{wang2016statistical}
\bibinfo{author}{\bibfnamefont{Z.}~\bibnamefont{Wang}},
  \bibinfo{author}{\bibfnamefont{C.~T.} \bibnamefont{Bauch}},
  \bibinfo{author}{\bibfnamefont{S.}~\bibnamefont{Bhattacharyya}},
  \bibinfo{author}{\bibfnamefont{A.}~\bibnamefont{d'Onofrio}},
  \bibinfo{author}{\bibfnamefont{P.}~\bibnamefont{Manfredi}},
  \bibinfo{author}{\bibfnamefont{M.}~\bibnamefont{Perc}},
  \bibinfo{author}{\bibfnamefont{N.}~\bibnamefont{Perra}},
  \bibinfo{author}{\bibfnamefont{M.}~\bibnamefont{Salath{\'e}}},
  \bibnamefont{and} \bibinfo{author}{\bibfnamefont{D.}~\bibnamefont{Zhao}},
  \bibinfo{journal}{Physics Reports} \textbf{\bibinfo{volume}{664}},
  \bibinfo{pages}{1} (\bibinfo{year}{2016}).

\bibitem[{\citenamefont{Takaguchi et~al.}(2014)\citenamefont{Takaguchi,
  Hasegawa, and Yoshida}}]{takaguchi2014suppressing}
\bibinfo{author}{\bibfnamefont{T.}~\bibnamefont{Takaguchi}},
  \bibinfo{author}{\bibfnamefont{T.}~\bibnamefont{Hasegawa}}, \bibnamefont{and}
  \bibinfo{author}{\bibfnamefont{Y.}~\bibnamefont{Yoshida}},
  \bibinfo{journal}{Physical Review E} \textbf{\bibinfo{volume}{90}},
  \bibinfo{pages}{012807} (\bibinfo{year}{2014}).

\bibitem[{\citenamefont{Hasegawa and Nemoto}(2017)}]{hasegawa2017efficiency}
\bibinfo{author}{\bibfnamefont{T.}~\bibnamefont{Hasegawa}} \bibnamefont{and}
  \bibinfo{author}{\bibfnamefont{K.}~\bibnamefont{Nemoto}},
  \bibinfo{journal}{Physical Review E} \textbf{\bibinfo{volume}{96}},
  \bibinfo{pages}{022311} (\bibinfo{year}{2017}).

\bibitem[{\citenamefont{Strona and Castellano}(2018)}]{strona2018rapid}
\bibinfo{author}{\bibfnamefont{G.}~\bibnamefont{Strona}} \bibnamefont{and}
  \bibinfo{author}{\bibfnamefont{C.}~\bibnamefont{Castellano}},
  \bibinfo{journal}{Physical Review E} \textbf{\bibinfo{volume}{97}},
  \bibinfo{pages}{022308} (\bibinfo{year}{2018}).

\bibitem[{\citenamefont{Gutin et~al.}(2021)\citenamefont{Gutin, Hirano, Hwang,
  Neary, and Toda}}]{gutin2021effect}
\bibinfo{author}{\bibfnamefont{G.}~\bibnamefont{Gutin}},
  \bibinfo{author}{\bibfnamefont{T.}~\bibnamefont{Hirano}},
  \bibinfo{author}{\bibfnamefont{S.-H.} \bibnamefont{Hwang}},
  \bibinfo{author}{\bibfnamefont{P.~R.} \bibnamefont{Neary}}, \bibnamefont{and}
  \bibinfo{author}{\bibfnamefont{A.~A.} \bibnamefont{Toda}},
  \bibinfo{journal}{Journal of Economic Interaction and Coordination} pp.
  \bibinfo{pages}{1--19} (\bibinfo{year}{2021}).

\bibitem[{\citenamefont{Maharaj and
  Kleczkowski}(2012)}]{maharaj2012controlling}
\bibinfo{author}{\bibfnamefont{S.}~\bibnamefont{Maharaj}} \bibnamefont{and}
  \bibinfo{author}{\bibfnamefont{A.}~\bibnamefont{Kleczkowski}},
  \bibinfo{journal}{BMC Public Health} \textbf{\bibinfo{volume}{12}},
  \bibinfo{pages}{1} (\bibinfo{year}{2012}).

\bibitem[{\citenamefont{Valdez et~al.}(2012)\citenamefont{Valdez, Macri, and
  Braunstein}}]{valdez2012intermittent}
\bibinfo{author}{\bibfnamefont{L.}~\bibnamefont{Valdez}},
  \bibinfo{author}{\bibfnamefont{P.~A.} \bibnamefont{Macri}}, \bibnamefont{and}
  \bibinfo{author}{\bibfnamefont{L.~A.} \bibnamefont{Braunstein}},
  \bibinfo{journal}{Physical Review E} \textbf{\bibinfo{volume}{85}},
  \bibinfo{pages}{036108} (\bibinfo{year}{2012}).

\bibitem[{\citenamefont{Maheshwari and Albert}(2020)}]{maheshwari2020network}
\bibinfo{author}{\bibfnamefont{P.}~\bibnamefont{Maheshwari}} \bibnamefont{and}
  \bibinfo{author}{\bibfnamefont{R.}~\bibnamefont{Albert}},
  \bibinfo{journal}{Applied Network Science} \textbf{\bibinfo{volume}{5}},
  \bibinfo{pages}{1} (\bibinfo{year}{2020}).

\bibitem[{\citenamefont{Nande et~al.}(2021)\citenamefont{Nande, Adlam, Sheen,
  Levy, and Hill}}]{nande2021dynamics}
\bibinfo{author}{\bibfnamefont{A.}~\bibnamefont{Nande}},
  \bibinfo{author}{\bibfnamefont{B.}~\bibnamefont{Adlam}},
  \bibinfo{author}{\bibfnamefont{J.}~\bibnamefont{Sheen}},
  \bibinfo{author}{\bibfnamefont{M.~Z.} \bibnamefont{Levy}}, \bibnamefont{and}
  \bibinfo{author}{\bibfnamefont{A.~L.} \bibnamefont{Hill}},
  \bibinfo{journal}{PLoS Computational Biology} \textbf{\bibinfo{volume}{17}},
  \bibinfo{pages}{e1008684} (\bibinfo{year}{2021}).

\bibitem[{\citenamefont{Gosak et~al.}(2021)\citenamefont{Gosak, Duh,
  Markovi{\v{c}}, and Perc}}]{gosak2021community}
\bibinfo{author}{\bibfnamefont{M.}~\bibnamefont{Gosak}},
  \bibinfo{author}{\bibfnamefont{M.}~\bibnamefont{Duh}},
  \bibinfo{author}{\bibfnamefont{R.}~\bibnamefont{Markovi{\v{c}}}},
  \bibnamefont{and} \bibinfo{author}{\bibfnamefont{M.}~\bibnamefont{Perc}},
  \bibinfo{journal}{New Journal of Physics} \textbf{\bibinfo{volume}{23}},
  \bibinfo{pages}{043039} (\bibinfo{year}{2021}).

\bibitem[{\citenamefont{Li et~al.}(2020)\citenamefont{Li, Liu, Li, Qian, and
  Dai}}]{li2020mask}
\bibinfo{author}{\bibfnamefont{T.}~\bibnamefont{Li}},
  \bibinfo{author}{\bibfnamefont{Y.}~\bibnamefont{Liu}},
  \bibinfo{author}{\bibfnamefont{M.}~\bibnamefont{Li}},
  \bibinfo{author}{\bibfnamefont{X.}~\bibnamefont{Qian}}, \bibnamefont{and}
  \bibinfo{author}{\bibfnamefont{S.~Y.} \bibnamefont{Dai}},
  \bibinfo{journal}{PloS One} \textbf{\bibinfo{volume}{15}},
  \bibinfo{pages}{e0237691} (\bibinfo{year}{2020}).

\bibitem[{\citenamefont{Eikenberry et~al.}(2020)\citenamefont{Eikenberry,
  Mancuso, Iboi, Phan, Eikenberry, Kuang, Kostelich, and
  Gumel}}]{eikenberry2020mask}
\bibinfo{author}{\bibfnamefont{S.~E.} \bibnamefont{Eikenberry}},
  \bibinfo{author}{\bibfnamefont{M.}~\bibnamefont{Mancuso}},
  \bibinfo{author}{\bibfnamefont{E.}~\bibnamefont{Iboi}},
  \bibinfo{author}{\bibfnamefont{T.}~\bibnamefont{Phan}},
  \bibinfo{author}{\bibfnamefont{K.}~\bibnamefont{Eikenberry}},
  \bibinfo{author}{\bibfnamefont{Y.}~\bibnamefont{Kuang}},
  \bibinfo{author}{\bibfnamefont{E.}~\bibnamefont{Kostelich}},
  \bibnamefont{and} \bibinfo{author}{\bibfnamefont{A.~B.} \bibnamefont{Gumel}},
  \bibinfo{journal}{Infectious Disease Modelling} \textbf{\bibinfo{volume}{5}},
  \bibinfo{pages}{293} (\bibinfo{year}{2020}).

\bibitem[{\citenamefont{Leung et~al.}(2020)\citenamefont{Leung, Ding, Huang,
  and Rabbany}}]{leung2020contact}
\bibinfo{author}{\bibfnamefont{A.}~\bibnamefont{Leung}},
  \bibinfo{author}{\bibfnamefont{X.}~\bibnamefont{Ding}},
  \bibinfo{author}{\bibfnamefont{S.}~\bibnamefont{Huang}}, \bibnamefont{and}
  \bibinfo{author}{\bibfnamefont{R.}~\bibnamefont{Rabbany}},
  \bibinfo{journal}{arXiv preprint arXiv:2010.03081}  (\bibinfo{year}{2020}).

\bibitem[{\citenamefont{Ya{\u{g}}an et~al.}(2021)\citenamefont{Ya{\u{g}}an,
  Sridhar, Eletreby, Levin, B.~Plotkin, and Poor}}]{yagan2021modeling}
\bibinfo{author}{\bibfnamefont{O.}~\bibnamefont{Ya{\u{g}}an}},
  \bibinfo{author}{\bibfnamefont{A.}~\bibnamefont{Sridhar}},
  \bibinfo{author}{\bibfnamefont{R.}~\bibnamefont{Eletreby}},
  \bibinfo{author}{\bibfnamefont{S.~A.} \bibnamefont{Levin}},
  \bibinfo{author}{\bibfnamefont{J.}~\bibnamefont{B.~Plotkin}},
  \bibnamefont{and} \bibinfo{author}{\bibfnamefont{H.~V.} \bibnamefont{Poor}},
  \bibinfo{journal}{Harvard Data Science Review}  (\bibinfo{year}{2021}),
  \urlprefix\url{https://hdsr.mitpress.mit.edu/pub/2q9jiymv}.

\bibitem[{\citenamefont{Tian et~al.}(2021{\natexlab{a}})\citenamefont{Tian,
  Sridhar, Ya{\u{g}}an, and Poor}}]{tian2021analysis}
\bibinfo{author}{\bibfnamefont{Y.}~\bibnamefont{Tian}},
  \bibinfo{author}{\bibfnamefont{A.}~\bibnamefont{Sridhar}},
  \bibinfo{author}{\bibfnamefont{O.}~\bibnamefont{Ya{\u{g}}an}},
  \bibnamefont{and} \bibinfo{author}{\bibfnamefont{H.~V.} \bibnamefont{Poor}},
  in \emph{\bibinfo{booktitle}{2021 American Control Conference (ACC)}}
  (\bibinfo{year}{2021}{\natexlab{a}}), pp. \bibinfo{pages}{3132--3137}.

\bibitem[{\citenamefont{Ying and O'Clery}(2021)}]{ying2021modelling}
\bibinfo{author}{\bibfnamefont{F.}~\bibnamefont{Ying}} \bibnamefont{and}
  \bibinfo{author}{\bibfnamefont{N.}~\bibnamefont{O'Clery}},
  \bibinfo{journal}{Plos One} \textbf{\bibinfo{volume}{16}},
  \bibinfo{pages}{e0249821} (\bibinfo{year}{2021}).

\bibitem[{\citenamefont{Sridhar et~al.}(2021)\citenamefont{Sridhar,
  Ya{\u{g}}an, Eletreby, Levin, Plotkin, and Poor}}]{sridhar2021leveraging}
\bibinfo{author}{\bibfnamefont{A.}~\bibnamefont{Sridhar}},
  \bibinfo{author}{\bibfnamefont{O.}~\bibnamefont{Ya{\u{g}}an}},
  \bibinfo{author}{\bibfnamefont{R.}~\bibnamefont{Eletreby}},
  \bibinfo{author}{\bibfnamefont{S.~A.} \bibnamefont{Levin}},
  \bibinfo{author}{\bibfnamefont{J.~B.} \bibnamefont{Plotkin}},
  \bibnamefont{and} \bibinfo{author}{\bibfnamefont{H.~V.} \bibnamefont{Poor}},
  in \emph{\bibinfo{booktitle}{ICASSP 2021-2021 IEEE International Conference
  on Acoustics, Speech and Signal Processing (ICASSP)}} (\bibinfo{year}{2021}),
  pp. \bibinfo{pages}{8163--8167}.

\bibitem[{\citenamefont{Lee and Zhu}(2021)}]{lee2021epidemic}
\bibinfo{author}{\bibfnamefont{D.-S.} \bibnamefont{Lee}} \bibnamefont{and}
  \bibinfo{author}{\bibfnamefont{M.}~\bibnamefont{Zhu}}, \bibinfo{journal}{IEEE
  Transactions on Computational Social Systems} pp. \bibinfo{pages}{1--14}
  (\bibinfo{year}{2021}).

\bibitem[{\citenamefont{Tian et~al.}(2021{\natexlab{b}})\citenamefont{Tian,
  Sridhar, Poor, and Ya{\u{g}}an}}]{tian2021role}
\bibinfo{author}{\bibfnamefont{Y.}~\bibnamefont{Tian}},
  \bibinfo{author}{\bibfnamefont{A.}~\bibnamefont{Sridhar}},
  \bibinfo{author}{\bibfnamefont{H.~V.} \bibnamefont{Poor}}, \bibnamefont{and}
  \bibinfo{author}{\bibfnamefont{O.}~\bibnamefont{Ya{\u{g}}an}},
  \bibinfo{journal}{arXiv preprint arXiv:2110.04398}
  (\bibinfo{year}{2021}{\natexlab{b}}).

\bibitem[{\citenamefont{Alexander and Day}(2010)}]{alexander2010risk}
\bibinfo{author}{\bibfnamefont{H.}~\bibnamefont{Alexander}} \bibnamefont{and}
  \bibinfo{author}{\bibfnamefont{T.}~\bibnamefont{Day}},
  \bibinfo{journal}{Journal of The Royal Society Interface}
  \textbf{\bibinfo{volume}{7}}, \bibinfo{pages}{1455} (\bibinfo{year}{2010}).

\bibitem[{\citenamefont{Eletreby et~al.}(2020)\citenamefont{Eletreby, Zhuang,
  Carley, Ya{\u{g}}an, and Poor}}]{eletreby2020effects}
\bibinfo{author}{\bibfnamefont{R.}~\bibnamefont{Eletreby}},
  \bibinfo{author}{\bibfnamefont{Y.}~\bibnamefont{Zhuang}},
  \bibinfo{author}{\bibfnamefont{K.~M.} \bibnamefont{Carley}},
  \bibinfo{author}{\bibfnamefont{O.}~\bibnamefont{Ya{\u{g}}an}},
  \bibnamefont{and} \bibinfo{author}{\bibfnamefont{H.~V.} \bibnamefont{Poor}},
  \bibinfo{journal}{Proceedings of the National Academy of Sciences}
  \textbf{\bibinfo{volume}{117}}, \bibinfo{pages}{5664} (\bibinfo{year}{2020}).

\bibitem[{\citenamefont{McPherson et~al.}(2001)\citenamefont{McPherson,
  Smith-Lovin, and Cook}}]{mcpherson2001birds}
\bibinfo{author}{\bibfnamefont{M.}~\bibnamefont{McPherson}},
  \bibinfo{author}{\bibfnamefont{L.}~\bibnamefont{Smith-Lovin}},
  \bibnamefont{and} \bibinfo{author}{\bibfnamefont{J.~M.} \bibnamefont{Cook}},
  \bibinfo{journal}{Annual Review of Sociology} \textbf{\bibinfo{volume}{27}},
  \bibinfo{pages}{415} (\bibinfo{year}{2001}).

\bibitem[{\citenamefont{Centola}(2011)}]{centola2011experimental}
\bibinfo{author}{\bibfnamefont{D.}~\bibnamefont{Centola}},
  \bibinfo{journal}{Science} \textbf{\bibinfo{volume}{334}},
  \bibinfo{pages}{1269} (\bibinfo{year}{2011}).

\bibitem[{\citenamefont{Haischer et~al.}(2020)\citenamefont{Haischer, Beilfuss,
  Hart, Opielinski, Wrucke, Zirgaitis, Uhrich, and
  Hunter}}]{haischer2020wearing}
\bibinfo{author}{\bibfnamefont{M.~H.} \bibnamefont{Haischer}},
  \bibinfo{author}{\bibfnamefont{R.}~\bibnamefont{Beilfuss}},
  \bibinfo{author}{\bibfnamefont{M.~R.} \bibnamefont{Hart}},
  \bibinfo{author}{\bibfnamefont{L.}~\bibnamefont{Opielinski}},
  \bibinfo{author}{\bibfnamefont{D.}~\bibnamefont{Wrucke}},
  \bibinfo{author}{\bibfnamefont{G.}~\bibnamefont{Zirgaitis}},
  \bibinfo{author}{\bibfnamefont{T.~D.} \bibnamefont{Uhrich}},
  \bibnamefont{and} \bibinfo{author}{\bibfnamefont{S.~K.}
  \bibnamefont{Hunter}}, \bibinfo{journal}{PloS One}
  \textbf{\bibinfo{volume}{15}}, \bibinfo{pages}{e0240785}
  (\bibinfo{year}{2020}).

\bibitem[{\citenamefont{Woodcock and Schultz}(2021)}]{woodcock2021role}
\bibinfo{author}{\bibfnamefont{A.}~\bibnamefont{Woodcock}} \bibnamefont{and}
  \bibinfo{author}{\bibfnamefont{P.~W.} \bibnamefont{Schultz}},
  \bibinfo{journal}{PloS One} \textbf{\bibinfo{volume}{16}},
  \bibinfo{pages}{e0261321} (\bibinfo{year}{2021}).

\bibitem[{\citenamefont{Newman}(2003)}]{newman2003mixing}
\bibinfo{author}{\bibfnamefont{M.~E.~J.} \bibnamefont{Newman}},
  \bibinfo{journal}{Physical Review E} \textbf{\bibinfo{volume}{67}},
  \bibinfo{pages}{026126} (\bibinfo{year}{2003}).

\bibitem[{\citenamefont{Newman}(2002)}]{newman2002spread}
\bibinfo{author}{\bibfnamefont{M.~E.~J.} \bibnamefont{Newman}},
  \bibinfo{journal}{Physical Review E} \textbf{\bibinfo{volume}{66}},
  \bibinfo{pages}{016128} (\bibinfo{year}{2002}).

\bibitem[{\citenamefont{Brienen et~al.}(2010)\citenamefont{Brienen, Timen,
  Wallinga, Van~Steenbergen, and Teunis}}]{brienen2010effect}
\bibinfo{author}{\bibfnamefont{N.~C.} \bibnamefont{Brienen}},
  \bibinfo{author}{\bibfnamefont{A.}~\bibnamefont{Timen}},
  \bibinfo{author}{\bibfnamefont{J.}~\bibnamefont{Wallinga}},
  \bibinfo{author}{\bibfnamefont{J.~E.} \bibnamefont{Van~Steenbergen}},
  \bibnamefont{and} \bibinfo{author}{\bibfnamefont{P.~F.}
  \bibnamefont{Teunis}}, \bibinfo{journal}{Risk Analysis: An International
  Journal} \textbf{\bibinfo{volume}{30}}, \bibinfo{pages}{1210}
  (\bibinfo{year}{2010}).

\bibitem[{\citenamefont{Lai et~al.}(2012)\citenamefont{Lai, Poon, and
  Cheung}}]{lai2012effectiveness}
\bibinfo{author}{\bibfnamefont{A.}~\bibnamefont{Lai}},
  \bibinfo{author}{\bibfnamefont{C.}~\bibnamefont{Poon}}, \bibnamefont{and}
  \bibinfo{author}{\bibfnamefont{A.}~\bibnamefont{Cheung}},
  \bibinfo{journal}{Journal of the Royal Society Interface}
  \textbf{\bibinfo{volume}{9}}, \bibinfo{pages}{938} (\bibinfo{year}{2012}).

\bibitem[{\citenamefont{Davies et~al.}(2013)\citenamefont{Davies, Thompson,
  Giri, Kafatos, Walker, and Bennett}}]{davies2013testing}
\bibinfo{author}{\bibfnamefont{A.}~\bibnamefont{Davies}},
  \bibinfo{author}{\bibfnamefont{K.-A.} \bibnamefont{Thompson}},
  \bibinfo{author}{\bibfnamefont{K.}~\bibnamefont{Giri}},
  \bibinfo{author}{\bibfnamefont{G.}~\bibnamefont{Kafatos}},
  \bibinfo{author}{\bibfnamefont{J.}~\bibnamefont{Walker}}, \bibnamefont{and}
  \bibinfo{author}{\bibfnamefont{A.}~\bibnamefont{Bennett}},
  \bibinfo{journal}{Disaster medicine and public health preparedness}
  \textbf{\bibinfo{volume}{7}}, \bibinfo{pages}{413} (\bibinfo{year}{2013}).

\bibitem[{\citenamefont{Ueki et~al.}(2020)\citenamefont{Ueki, Furusawa,
  Iwatsuki-Horimoto, Imai, Kabata, Nishimura, and
  Kawaoka}}]{ueki2020effectiveness}
\bibinfo{author}{\bibfnamefont{H.}~\bibnamefont{Ueki}},
  \bibinfo{author}{\bibfnamefont{Y.}~\bibnamefont{Furusawa}},
  \bibinfo{author}{\bibfnamefont{K.}~\bibnamefont{Iwatsuki-Horimoto}},
  \bibinfo{author}{\bibfnamefont{M.}~\bibnamefont{Imai}},
  \bibinfo{author}{\bibfnamefont{H.}~\bibnamefont{Kabata}},
  \bibinfo{author}{\bibfnamefont{H.}~\bibnamefont{Nishimura}},
  \bibnamefont{and} \bibinfo{author}{\bibfnamefont{Y.}~\bibnamefont{Kawaoka}},
  \bibinfo{journal}{MSphere} \textbf{\bibinfo{volume}{5}}
  (\bibinfo{year}{2020}).

\bibitem[{\citenamefont{Allard et~al.}(2009)\citenamefont{Allard, No{\"e}l,
  Dub{\'e}, and Pourbohloul}}]{allard2009heterogeneous}
\bibinfo{author}{\bibfnamefont{A.}~\bibnamefont{Allard}},
  \bibinfo{author}{\bibfnamefont{P.-A.} \bibnamefont{No{\"e}l}},
  \bibinfo{author}{\bibfnamefont{L.~J.} \bibnamefont{Dub{\'e}}},
  \bibnamefont{and}
  \bibinfo{author}{\bibfnamefont{B.}~\bibnamefont{Pourbohloul}},
  \bibinfo{journal}{Physical Review E} \textbf{\bibinfo{volume}{79}},
  \bibinfo{pages}{036113} (\bibinfo{year}{2009}).

\bibitem[{\citenamefont{Meyers et~al.}(2006)\citenamefont{Meyers, Newman, and
  Pourbohloul}}]{meyers2006predicting}
\bibinfo{author}{\bibfnamefont{L.~A.} \bibnamefont{Meyers}},
  \bibinfo{author}{\bibfnamefont{M.}~\bibnamefont{Newman}}, \bibnamefont{and}
  \bibinfo{author}{\bibfnamefont{B.}~\bibnamefont{Pourbohloul}},
  \bibinfo{journal}{Journal of theoretical biology}
  \textbf{\bibinfo{volume}{240}}, \bibinfo{pages}{400} (\bibinfo{year}{2006}).

\bibitem[{\citenamefont{Kenah and Robins}(2007)}]{kenah2007second}
\bibinfo{author}{\bibfnamefont{E.}~\bibnamefont{Kenah}} \bibnamefont{and}
  \bibinfo{author}{\bibfnamefont{J.~M.} \bibnamefont{Robins}},
  \bibinfo{journal}{Physical Review E} \textbf{\bibinfo{volume}{76}},
  \bibinfo{pages}{036113} (\bibinfo{year}{2007}).

\bibitem[{\citenamefont{Miller}(2007)}]{miller2007epidemic}
\bibinfo{author}{\bibfnamefont{J.~C.} \bibnamefont{Miller}},
  \bibinfo{journal}{Physical Review E} \textbf{\bibinfo{volume}{76}},
  \bibinfo{pages}{010101} (\bibinfo{year}{2007}).

\bibitem[{\citenamefont{Stauffer and Aharony}(2018)}]{stauffer2018introduction}
\bibinfo{author}{\bibfnamefont{D.}~\bibnamefont{Stauffer}} \bibnamefont{and}
  \bibinfo{author}{\bibfnamefont{A.}~\bibnamefont{Aharony}},
  \emph{\bibinfo{title}{Introduction to percolation theory}}
  (\bibinfo{publisher}{Taylor \& Francis}, \bibinfo{year}{2018}).

\bibitem[{\citenamefont{Grassberger}(1983)}]{grassberger1983critical}
\bibinfo{author}{\bibfnamefont{P.}~\bibnamefont{Grassberger}},
  \bibinfo{journal}{Mathematical Biosciences} \textbf{\bibinfo{volume}{63}},
  \bibinfo{pages}{157} (\bibinfo{year}{1983}).

\bibitem[{\citenamefont{Tom{\'e} and Ziff}(2010)}]{tome2010critical}
\bibinfo{author}{\bibfnamefont{T.}~\bibnamefont{Tom{\'e}}} \bibnamefont{and}
  \bibinfo{author}{\bibfnamefont{R.~M.} \bibnamefont{Ziff}},
  \bibinfo{journal}{Physical Review E} \textbf{\bibinfo{volume}{82}},
  \bibinfo{pages}{051921} (\bibinfo{year}{2010}).

\bibitem[{\citenamefont{De~Souza et~al.}(2011)\citenamefont{De~Souza, Tom{\'e},
  and Ziff}}]{de2011new}
\bibinfo{author}{\bibfnamefont{D.~R.} \bibnamefont{De~Souza}},
  \bibinfo{author}{\bibfnamefont{T.}~\bibnamefont{Tom{\'e}}}, \bibnamefont{and}
  \bibinfo{author}{\bibfnamefont{R.~M.} \bibnamefont{Ziff}},
  \bibinfo{journal}{Journal of Statistical Mechanics: Theory and Experiment}
  \textbf{\bibinfo{volume}{2011}}, \bibinfo{pages}{P03006}
  (\bibinfo{year}{2011}).

\bibitem[{\citenamefont{Duh et~al.}(2019)\citenamefont{Duh, Gosak, Slavinec,
  and Perc}}]{duh2019assortativity}
\bibinfo{author}{\bibfnamefont{M.}~\bibnamefont{Duh}},
  \bibinfo{author}{\bibfnamefont{M.}~\bibnamefont{Gosak}},
  \bibinfo{author}{\bibfnamefont{M.}~\bibnamefont{Slavinec}}, \bibnamefont{and}
  \bibinfo{author}{\bibfnamefont{M.}~\bibnamefont{Perc}}, \bibinfo{journal}{New
  Journal of Physics} \textbf{\bibinfo{volume}{21}}, \bibinfo{pages}{123016}
  (\bibinfo{year}{2019}).

\bibitem[{\citenamefont{Boccaletti et~al.}(2014)\citenamefont{Boccaletti,
  Bianconi, Criado, Del~Genio, G{\'o}mez-Gardenes, Romance, Sendina-Nadal,
  Wang, and Zanin}}]{boccaletti2014structure}
\bibinfo{author}{\bibfnamefont{S.}~\bibnamefont{Boccaletti}},
  \bibinfo{author}{\bibfnamefont{G.}~\bibnamefont{Bianconi}},
  \bibinfo{author}{\bibfnamefont{R.}~\bibnamefont{Criado}},
  \bibinfo{author}{\bibfnamefont{C.~I.} \bibnamefont{Del~Genio}},
  \bibinfo{author}{\bibfnamefont{J.}~\bibnamefont{G{\'o}mez-Gardenes}},
  \bibinfo{author}{\bibfnamefont{M.}~\bibnamefont{Romance}},
  \bibinfo{author}{\bibfnamefont{I.}~\bibnamefont{Sendina-Nadal}},
  \bibinfo{author}{\bibfnamefont{Z.}~\bibnamefont{Wang}}, \bibnamefont{and}
  \bibinfo{author}{\bibfnamefont{M.}~\bibnamefont{Zanin}},
  \bibinfo{journal}{Physics Reports} \textbf{\bibinfo{volume}{544}},
  \bibinfo{pages}{1} (\bibinfo{year}{2014}).

\bibitem[{\citenamefont{Bianconi}(2018)}]{bianconi2018multilayer}
\bibinfo{author}{\bibfnamefont{G.}~\bibnamefont{Bianconi}},
  \emph{\bibinfo{title}{Multilayer networks: structure and function}}
  (\bibinfo{publisher}{Oxford University Press}, \bibinfo{year}{2018}).

\bibitem[{\citenamefont{Torres and Bianconi}(2020)}]{torres2020simplicial}
\bibinfo{author}{\bibfnamefont{J.~J.} \bibnamefont{Torres}} \bibnamefont{and}
  \bibinfo{author}{\bibfnamefont{G.}~\bibnamefont{Bianconi}},
  \bibinfo{journal}{Journal of Physics: Complexity}
  \textbf{\bibinfo{volume}{1}}, \bibinfo{pages}{015002} (\bibinfo{year}{2020}).

\bibitem[{\citenamefont{Bianconi}(2021)}]{bianconi2021higher}
\bibinfo{author}{\bibfnamefont{G.}~\bibnamefont{Bianconi}},
  \emph{\bibinfo{title}{Higher Order Networks: An Introduction to Simplicial
  Complexes}} (\bibinfo{publisher}{Cambridge University Press},
  \bibinfo{year}{2021}).

\bibitem[{\citenamefont{Kryven}(2017)}]{kryven2017finite}
\bibinfo{author}{\bibfnamefont{I.}~\bibnamefont{Kryven}},
  \bibinfo{journal}{Physical Review E} \textbf{\bibinfo{volume}{96}},
  \bibinfo{pages}{052304} (\bibinfo{year}{2017}).

\bibitem[{\citenamefont{Kryven}(2019)}]{kryven2019bond}
\bibinfo{author}{\bibfnamefont{I.}~\bibnamefont{Kryven}},
  \bibinfo{journal}{Nature Communications} \textbf{\bibinfo{volume}{10}},
  \bibinfo{pages}{1} (\bibinfo{year}{2019}).

\bibitem[{\citenamefont{Bianconi et~al.}(2019)\citenamefont{Bianconi, Kryven,
  and Ziff}}]{bianconi2019percolation}
\bibinfo{author}{\bibfnamefont{G.}~\bibnamefont{Bianconi}},
  \bibinfo{author}{\bibfnamefont{I.}~\bibnamefont{Kryven}}, \bibnamefont{and}
  \bibinfo{author}{\bibfnamefont{R.~M.} \bibnamefont{Ziff}},
  \bibinfo{journal}{Physical Review E} \textbf{\bibinfo{volume}{100}},
  \bibinfo{pages}{062311} (\bibinfo{year}{2019}).

\bibitem[{\citenamefont{Balogh et~al.}(2020)\citenamefont{Balogh, Palla, and
  Kryven}}]{balogh2020networks}
\bibinfo{author}{\bibfnamefont{S.~G.} \bibnamefont{Balogh}},
  \bibinfo{author}{\bibfnamefont{G.}~\bibnamefont{Palla}}, \bibnamefont{and}
  \bibinfo{author}{\bibfnamefont{I.}~\bibnamefont{Kryven}},
  \bibinfo{journal}{Journal of Complex Networks} \textbf{\bibinfo{volume}{8}},
  \bibinfo{pages}{cnaa045} (\bibinfo{year}{2020}).

\bibitem[{\citenamefont{Hiraoka et~al.}(2022)\citenamefont{Hiraoka, Rizi,
  Kivel\"a, and Saram\"aki}}]{hiraoka2021herd}
\bibinfo{author}{\bibfnamefont{T.}~\bibnamefont{Hiraoka}},
  \bibinfo{author}{\bibfnamefont{A.~K.} \bibnamefont{Rizi}},
  \bibinfo{author}{\bibfnamefont{M.}~\bibnamefont{Kivel\"a}}, \bibnamefont{and}
  \bibinfo{author}{\bibfnamefont{J.}~\bibnamefont{Saram\"aki}},
  \bibinfo{journal}{Physical Review E} \textbf{\bibinfo{volume}{105}},
  \bibinfo{pages}{L052301} (\bibinfo{year}{2022}).

\bibitem[{\citenamefont{Burgio et~al.}(2022)\citenamefont{Burgio, Steinegger,
  and Arenas}}]{burgio2022homophily}
\bibinfo{author}{\bibfnamefont{G.}~\bibnamefont{Burgio}},
  \bibinfo{author}{\bibfnamefont{B.}~\bibnamefont{Steinegger}},
  \bibnamefont{and} \bibinfo{author}{\bibfnamefont{A.}~\bibnamefont{Arenas}},
  \bibinfo{journal}{Communications Physics} \textbf{\bibinfo{volume}{5}},
  \bibinfo{pages}{1} (\bibinfo{year}{2022}).

\end{thebibliography}
\end{document}